\documentclass[fleqn,usenatbib,useAMS]{mnras}

\usepackage{graphicx}	% Including figure files
\usepackage{amsmath}	% Advanced maths commands
\usepackage{amssymb}	% Extra maths symbols
\usepackage{multicol}        % Multi-column entries in tables
\usepackage{bm}		% Bold maths symbols, including upright Greek
\usepackage{pdflscape}	% Landscape pages
\usepackage{xcolor}
\usepackage{verbatim}
\usepackage[position=top]{subfig}

\newcommand{\Mpc}{\,$h^{-1}$Mpc\,} % 
\newcommand{\iMpc}{\,$h \text{Mpc}^{-1}$} % 
\def\[{\begin{equation}}
\def\]{\end{equation}}

\usepackage[T1]{fontenc}
\usepackage{ae,aecompl}

\usepackage{newtxtext,newtxmath}
\title[Projected two- and three-point statistics]{Projected two- and three-point statistics: Forecasts and mitigation of non-linear RSDs}

\author[O. Leicht et al.]{
Oliver Leicht,\thanks{E-mail: ol248@cam.ac.uk}
Tobias Baldauf,
James Fergusson
and Paul Shellard
\\
Centre for Theoretical Cosmology, DAMTP, University of Cambridge, CB3 0WA Cambridge, United Kingdom\\
}

\date{Accepted XXX. Received YYY; in original form ZZZ}

\pubyear{2020}

\begin{document}
\label{firstpage}
\pagerange{\pageref{firstpage}--\pageref{lastpage}}
\maketitle


\begin{abstract}
The combination of two- and three-point clustering statistics of galaxies and the underlying matter distribution has the potential to break degeneracies between cosmological parameters and nuisance parameters and can lead to significantly tighter constraints on parameters describing the composition of the Universe and the dynamics of inflation. Here we investigate the relation between biases in the estimated parameters and inaccurate modelling of non-linear redshift-space distortions for the power spectrum and bispectrum of projected galaxy density fields and lensing convergence. Non-linear redshift-space distortions are one of the leading systematic uncertainties in galaxy clustering. Projections along the line of sight suppress radial modes and are thus allowing a trade-off between biases due to non-linear redshift-space distortions and statistical uncertainties. We investigate this bias-error trade-off for a CMASS-like survey with a varying number of redshift bins. Improved modelling of the non-linear redshift-space distortions allows the recovery  of more radial information when controlling for biases. Not modelling non-linear redshift space distortions inflates error bars for almost all parameters by $20\%$. The information loss for the amplitude of local non-Gaussianities is smaller, since it is best constrained from large scales. In addition, we show empirically that one can recover more than 99\% of the 3D power spectrum information if the depth of the tomographic bins is reduced to 10 \Mpc. 
\end{abstract} 


%%%%%%%%%%%%%%%%%%%%%%%%%%%%%%%%%%%%%%%%%%%%%%%%%%

%%%%%%%%%%%%%%%%% BODY OF PAPER %%%%%%%%%%%%%%%%%%

% The MNRAS class isn't designed to include a table of contents, but for this document one is useful.
% I therefore have to do some kludging to make it work without masses of blank space.
%\begingroup
%\let\clearpage\relax
%\tableofcontents
%\endgroup
%\newpage

%%%%%%%%%%%%%%%%%%%%%%%%%%%%%%%%%%%%%%%%%%%%%%%%%%
%%%%%%%%%%%%%%%% Introduction %%%%%%%%%%%%%%%%%%%%

\section{Introduction}
Our understanding of the Universe has been shaped by observations of the cosmic microwave background (CMB) and large-scale structure of the Universe (LSS) over the last thirty years~\citep{PlanckCollaboration2018PlanckParameters,Riess2016Hubble,Alam2017SDSSBao}. Future CMB lensing experiments are expected to refine this picture~\citep{TheSimonsObservatoryCollaboration2018TheForecasts,Abazajian2019CMB-S4Plan}. Upcoming galaxy surveys that trace the matter distribution of the LSS such as LSST~\citep{LSSTScienceCollaboration2009LSST2.0}, SPHEREx~\citep{Dore2014CosmologySurvey}, Euclid~\citep{Amendola2016CosmologySatellite} and DESI~\citep{DESICollaboration2016TheDesign} will contribute complementary information about the late time evolution of the Universe. Moreover, those surveys will achieve exquisitely small statistical errors due to the vast volumes they cover and high number density of tracers they resolve. Given their three-dimensional origin, upcoming LSS datasets are predicted to eventually contain more information about cosmological parameters than the CMB.

The early Universe's density distribution was very close to a Gaussian random field~\citep{PlanckCollaboration2019PlanckNon-Gaussianity} which is fully described by the two-point correlation function or its Fourier transform, the power spectrum. The subsequent non-linear evolution changed the matter distribution which manifests itself in a modification of the power spectrum on small scales and non-vanishing higher order correlation functions. The bispectrum, which is the Fourier transform of the three-point-correlation function, is known to contain most of the non-linear information on mildly non-linear scales. In addition, it allows us to break degeneracies between bias and amplitude parameters~\citep{Scoccimarro2000TheObservations, Sefusatti2006CosmologyBispectrum}.

CMB-lensing captures the integrated effect of matter onto CMB photons along their path from the surface of last scattering through the LSS to us. Accordingly, lensing spectra are described by projected spectra of the LSS. But there are also very valid reasons to study galaxy clustering statistics in projection. Firstly, tomographic surveys infer the redshift bins of objects and not their precise positions. Moreover, CMB lensing - galaxy clustering cross-correlations require a two-dimensional clustering analysis. Lastly, projections offer a way to suppress non-linear redshift-space distortions (RSDs).

RSDs are generated by galaxies' peculiar velocities parallel to the line of sight (LOS). The resulting Doppler redshift is degenerate with the redshift from the Hubble flow which is used to determine the radial positions. On large scales, the effect is well described perturbatively, but on smaller scales one has to resort to empirical models. Work has been done in this direction~\citep{Scoccimarro1999TheSpace,Taruya2010TNSmodel, Gil-Marin2014DarkApplications,Slepian2017} but the fundamental issue of potentially biased estimates caused by inaccurate modelling remains. 

In this work we are quantifying the parameter shifts due to inaccurate RSD modelling. In particular, we study how those biases depend on the chosen projection depth and RSD model used. Given the exquisitely small statistical errors of upcoming surveys, it is worthwhile to make estimators more robust in order to confidently leverage the small statistical uncertainties. \cite{Agarwal2020} recently investigated the parameter shifts arising from an incomplete or incorrect account of bias parameters and selection effects.

In addition, we forecast error bars to investigate the relation between statistical and systematic uncertainties. The constraining power of the galaxy bispectrum for future galaxy surveys has been studied in~\cite{Yankelevich2019CosmologicalBispectrum, Karagiannis2018ConstrainingSurveys, Agarwal2020} and power spectrum forecasts for CMB-lensing - galaxy-clustering cross-correlations were performed in~\cite{Schmittfull2018ParameterClustering}. 

While we employ the flat sky approximation in this work, there is a growing literature that studies angular (cross-)correlation functions~\citep{Schoneberg2018BeyondStatistics, Gebhardt20172-FAST:Functions, Simonovic2017CosmologicalIntegrals, Slepian2018OnTheory, Fang2020BeyondLensing,Slepian2019RotationFunction, Campagne2017ADistortions, Tansella2018TheAspects, Tansella2018COFFE:Function,Campagne2017Angpow:Spectra, Dizgah2020CapturingSkew-Spectra}. The step from flat to curved sky is conceptually straightforward in our framework using the FFTlog-algorithm~\citep{Assassi2017EfficientStatistics}, but we leave this for future work.

This paper is structured as follows: We first introduce the power spectrum and bispectrum for matter and galaxy (cross-)correlations in section~\ref{sec:Theo3D}. We then project those into observable 2D spectra in section~\ref{sec:Projections}. In section~\ref{sec:Metho} we discuss the inference techniques used in our analysis. Our results are presented in section~\ref{sec:Results} and we conclude in section~\ref{sec:Conc}.

%%%%%%%%%%%%%%%%%%%%%%%%%%%%%%%%%%%%%%%%%%%%%%%%%%
%%%%%%%%%%%%% Statistics in 3D %%%%%%%%%%%%%%%%%%%

\section{Statistics in 3D}
\label{sec:Theo3D}
In this section we review the leading order power spectra and bispectra at late times. Starting with the matter predictions, we subsequently include galaxy biasing, RSDs and primordial non-Gaussianities (PNGs) into the model. We conclude this section with an overview of all matter-galaxy power spectra and bispectra.

\subsection{Matter power spectrum and bispectrum}
The linear matter power spectrum, $P_{\text{mm}}$, can be efficiently computed using Boltzmann codes such as \textsc{camb}~\citep{Lewis2000EFFICIENTMODELS,Challinor2011TheCounts}.\footnote{https://camb.info/} Using the matter power spectrum as input, the tree-level matter bispectrum is given by~\citep{Bernardeau2002Large-ScaleTheory}
\[ B_{\text{mmm} }(k_1, k_2, k_3) = 2 P_{\text{mm} }(k_1) P_{\text{mm} }(k_2) F_2(\bmath{k_1}, \bmath{k_2})+2\, \text{perm} . \, ,\]
where the second order gravitational kernel, $F_2$, is
\begin{equation}
    F_2(\bmath{k_1}, \bmath{k_2}) = \frac{5}{7} + \frac{1}{2} \mu \left(\frac{k_1}{k_2}+\frac{k_2}{k_1} \right) + \frac{2}{7}\mu^2, 
\end{equation}
with $\mu$ being the cosine of the angle between the two vectors $\bmath{k_1}$ and $\bmath{k_2}$. The second order kernel for the velocity divergence is given by
\begin{equation}
    G_2(\bmath{k_1}, \bmath{k_2}) = \frac{3}{7} + \frac{1}{2} \mu \left(\frac{k_1}{k_2}+\frac{k_2}{k_1} \right) + \frac{4}{7}\mu^2 .
\end{equation}
$G_2$ will be relevant for the perturbative description of RSDs.

\subsection{Biasing}
\label{ssec:Bias}
Galaxies do not directly trace the underlying matter distribution, leading to a so called bias relation between galaxy and the matter distribution. At large scales the bias relation can be described perturbatively, for a review see~\cite{Desjacques2018Large-scaleBias}.
Following~\cite{Mcdonald2009ClusteringLSS,Baldauf2012EvidenceBispectrum,Chan2012GravityBias, Lazeyras2017BeyondParameters, Abidi2018CubicSpace}, we express the galaxy over-density at late times as a function of the three bias parameters $(b_1, b_2, b_{\text{s}^2})$ together with some stochastic bias or shot-noise, $\epsilon$, caused by the discrete nature of galaxies
\begin{equation}
    \begin{split}
         \delta_g(\bmath{k}) &= b_1 \, \delta(\bmath{k})  + \epsilon(\bmath{k}) \\ &+ \frac{1}{2} \int \frac{d^3\bmath{q}}{(2 \pi)^3} \delta(\bmath{q}) \delta(\bmath{k-q})\big[b_2 + b_{s^2} S_2(\bmath{q},\bmath{k-q}) \big] \\
         &+\int \frac{d^3\bmath{q}}{(2 \pi)^3}\epsilon_\delta(\bmath{q}) \delta(\bmath{k-q}) \label{eq:bias} .
    \end{split}
\end{equation}
We truncated the expansion at second order, because we will be working with the tree-level power and bispectrum. The operator generating the Fourier representation of the square of the tidal tensor, $S_2$, is given by
\[ S_2(\bmath{k}, \bmath{q})  =  \frac{(\bmath{k} \cdot \bmath{q})^2}{(k q)^2} - \frac{1}{3}.\]
Following \cite{Scoccimarro2001PoissonShotNoise, Schmidt2016TowardsStructure,Desjacques2018Large-scaleBias}, we model the shot-noise as Poissonian, which leads to the following non-zero spectra
\[ P_{\epsilon \epsilon}= 1/\bar{n},  \qquad  P_{\epsilon \epsilon_\delta}= b_1/( 2 \bar{n}),  \qquad   B_{\epsilon \epsilon \epsilon} = 1/\bar{n}^2 \, ,\]
where $\bar{n}$ is the co-moving average number density of the tracers and all correlators between stochastic and matter densities fields are vanishing. Thus, the leading order galaxy power spectrum and bispectrum are given by
\[ P_{\text{gg}}(k) = b_1^2 P_{\text{mm} }(k) + \frac{1}{\bar{n}},\]
and
\begin{equation}
    \begin{split}
        B_{\text{ggg} }(k_1,k_2,k_3) = &\left[b_1^2 P_{\text{mm} }(k_1) P_{\text{mm} }(k_2) \right. \times \\ &\,\times \left(2 b_1 F_2(\bmath{k_1}, \bmath{k_2}) + b_2 + b_{s^2} S_2(\bmath{k_1}, \bmath{k_2})  \right) \\ &\,+ \left. \frac{b_1^2 }{\bar{n}} P_{\text{mm} }(k_1) +2\, \text{perm.} \right] + \frac{1}{\bar{n}^2}.  
    \end{split}{}
\end{equation}{} 

\subsection{Redshift space distortions}
Galaxies' radial distances are measured via their redshifts. However, peculiar velocities parallel to the LOS give rise to a Doppler redshift that is degenerate with the cosmological redshift and does bias distance measurements. 

On large scales, RSDs are caused by the coherent infall of galaxies into gravitational potentials and lead to an enhancement of modes parallel to the LOS. This effect can be treated perturbatively and incorporated into the spectra by means of redshift kernels $Z_i$ that replace the gravitational kernels $F_i$. The first two are given by~\cite{Scoccimarro1999TheSpace}
\begin{equation}
    Z_1(\bmath{k_i}) = b_1 + f \mu_i^2 , 
\end{equation}
where $f= d \ln D/ d \ln a$ refers to the logarithmic growth rate, the logarithmic derivative of the growth factor $D$. $\mu_i$ is the cosine of the angle between the wave vector and the LOS, i.e. $\mu_i=k_{i,\parallel}/k_i$.
\begin{equation}
\begin{split}
Z_2(\bmath{k_i},\bmath{k_j}) =& b_1 F_2(\bmath{k_i},\bmath{k_j}) +  \frac{b_2}{2} + \frac{b_{s^2}}{2} S_2(\bmath{k_i},\bmath{k_j}) \\ +&f \mu_{ij}^2 G_2(\bmath{k_i},\bmath{k_j})  \\ +& \frac{f \mu_{ij} k_{ij} }{2} \left[ \frac{\mu_i}{k_i} Z_1(\bmath{k_j}) + \frac{\mu_j}{k_j} Z_1(\bmath{k_i})  \right] .
\end{split}
\end{equation}
Here, $k_{ij}^2= (\bmath{k_i}+\bmath{k_j})^2$ and $\mu_{ij} k_{ij} = \mu_i k_i + \mu_j k_j$.\footnote{When projecting with a symmetric window function, the third line of the $Z_2$ kernel integrates to zero since $ \mu_{ij} k_{ij} = \mu_i k_i + \mu_j k_j  = -k_{l,\parallel}$ where 'l' labels the third vector.}
On small scales, the high and incoherent velocities of galaxies within potential wells, give rise to the Finger-of-God (FoG) effect, i.e., structures appear elongated along the LOS. This corresponds to a damping of small-scale modes along the LOS. As it is intrinsically non-linear, a perturbative treatment is not possible and phenomenological modelling is needed~\citep{Scoccimarro1999TheSpace}. For this work we follow~\cite{ Peacock1994ReconstructingFluctuations,Ballinger1996MeasuringSurveys} and use a Gaussian damping prefactor for the perturbative predictions for the power spectrum
\[ D_{\text{FoG} }^{\text{P}}(k) = \exp \left[-(k_{\parallel} \sigma_P)^2 \right] \, \label{eq:PS-FoG} ,\]
and a similar form for the bispectrum
\[ D_{\text{FoG}}^{\text{B}}(k_1, \dots, k_n) = \prod_{i=1}^n \exp \left[ -k_{i,\parallel}^2 \sigma_B^2/2 \right]. \label{eq:BS-FoG}\]
We treat $\sigma_P$ and $\sigma_B$ as free parameters whose values are inspired by $N$-body simulations. Physically speaking, they correspond to the velocity dispersion of the galaxy velocity distribution. The general definition of~(\ref{eq:BS-FoG}) will shorten notation later in the paper. 

\subsection{Primordial non-Gaussianities}
The assumption of Gaussian initial conditions can be tested by allowing small deviations and constraining their amplitude $f_{\text{NL}}$. This is typically done by adding primordial non-Gaussianities (PNGs) of known shape. Those give rise to two types of new terms: Firstly, the added PNGs lead to a non-zero matter bispectrum at all times. Secondly, the bias expansion~(\ref{eq:bias}) has to be carried out both in the density and gravitational potential adding new bias terms~\citep{Dalal2008ImprintsObjects,  McDonald2008PrimordialModel,Verde2009DetectabilityBias, Giannantonio2010StructureSimulations, BaldaufSenatore2011ngbias,Tellarini2015Non-localNon-Gaussianity, Assassi2015GalaxyNon-Gaussianity}. Since we use a fiducial cosmology without PNGs, only leading order (linear) terms in $f_{\text{NL}}$ are relevant for the Fisher forecasts and kept in the equations. 

Commonly the local, equilateral and orthogonal templates are used as proxies for more general PNGs. Those proxies are chosen to test particular aspects of inflation. The discovery of a non-zero PNG of the local shape
\[ B_{\phi}^{\text{loc} } = 2 f_{\text{NL}}^{\text{loc}} \left[ P_{\phi}(k_1)P_{\phi}(k_2) + 2\, \text{perms.}  \right] ,\]
 would be a strong indicator for multifield inflation~\citep{Creminelli2004AFunction}. A signal of the equilateral shape 
\begin{equation}
\begin{split}
B_{\phi}^{\text{equi} } = 6 f_{\text{NL}}^{\text{equi}}  &\biggl\{ - \left[  P_{\phi}(k_1)\,P_{\phi}(k_2) + 2\, \text{perms.}  \right] \biggr. \\ &  - 2 \left[ P_{\phi}(k_1)\, P_{\phi}(k_2)\, P_{\phi}(k_3) \right]^{2/3}  \\
& + \left. \left[ P_{\phi}^{1/3}(k_1)\, P_{\phi}^{2/3}(k_2)\, P_{\phi}(k_3) + 5\, \text{perms.} \right] \right\} ,
 \end{split}
\end{equation}
arises in a wide range of non-vanilla inflationary dynamics, for instance with non-standard kinetic terms, (see~\cite{2014A&A...571A..24P} and references therein). 
Lastly, the orthogonal shape, 
\begin{equation}
\begin{split}
B_{\phi}^{ \text{ortho} } = 6 f_{\text{NL}}^{\text{ortho} } &\left\{  -3 \left[  P_{\phi}(k_1)\,P_{\phi}(k_2) + 2\, \text{perms.}  \right] \right. \\ &- 8 \left[ P_{\phi}(k_1)\, P_{\phi}(k_2) \, P_{\phi}(k_3) \right]^{2/3} \\
+  3 &\left. \left[ P_{\phi}^{1/3}(k_1) \, P_{\phi}^{2/3}(k_2) \, P_{\phi}(k_3) + 5\, \text{perms.} \right] \right\} ,
 \end{split}
\end{equation}
 probes derivative interactions in multifield inflation~\citep{Senatore2010Non-GaussianitiesData}. 
 The mapping of the primordial gravitational potential $\phi$ to the late time density contrast $\delta$ is done using Poisson's equation together with the matter transfer function $T$ normalized to unity on large scales
\[ \delta = M(k,z) \, \phi \, , \] 
where the Poisson factor is
\[ M(k,z) = \frac{2 k^2 c^2 T(k) D(z)}{3 \Omega_{m,0} H_0^2 }\,, \]
with the Hubble constant, $H_0$, speed-of-light, $c$, growth factor $D$ and matter density $\Omega_{m,0}$. The linear matter power spectrum is obtained from the gravitational potential correlators at early times via
\[ P_{\text{mm} }(k) = M^2(k,z) P_{\phi}(k) . \]
A non-zero matter bispectrum at early times is proof of PNGs. The discussed proxies for PNGs are translated to late times via
\[ B_{\text{prim}}(k_1, k_2, k_3) = M(k_1,z)\, M(k_2,z)\, M(k_3,z)\, B_{\phi}(k_1, k_2,k_3) . \]
A complete bias expansion for isotropic and quadratic PNGs in Lagrangian space can be written using the field $\Psi$ that captures the non-Gaussianities in the primordial potential~\citep{Assassi2015GalaxyNon-Gaussianity}
\[\Psi(\bmath{k}) = A \int \frac{d^3 k_s}{(2 \pi)^3}\, \left(\frac{k}{k_s} \right)^\alpha \phi(\bmath{k_s}) \phi(\bmath{k-k_s}) \label{eq:Psi} . \]
In the squeezed limit, where the scale dependent bias terms are most relevant, the parameters $(A,\alpha)$ can be determined~\citep{Schmidt2010HaloNon-Gaussianity}. They are (1,0) for the local, (3,2) for the equilateral and (-3,1) for the orthogonal shape. Translating the expansion into Eulerian space, leads to the following additional terms in the bias expansion~\citep{Dalal2008ImprintsObjects, McDonald2008PrimordialModel, Verde2009DetectabilityBias,Giannantonio2010StructureSimulations, BaldaufSenatore2011ngbias,Assassi2015GalaxyNon-Gaussianity}
\begin{equation}
\begin{split}
&\delta_{\text{g}}^{(\text{NG})}(\bmath{k}) = b_\Psi \Psi(\bmath{k}) + \\ &\int\frac{d^3 \bmath{q}}{(2 \pi)^3}  \left[  \left(  b_{\Psi \delta} - b_\Psi N_2(\bmath{q},\bmath{k-q}) \right)\delta(\bmath{q}) + \epsilon_\Psi(\bmath{q}) \right] \Psi(\bmath{k-q}) .   \label{eq:NGbias}
\end{split}
\end{equation}
The $N_2$ kernel originates from the linear displacement field that maps Lagrangian to Eulerian space~(\ref{eq:Psi}) and is given by~\citep{Tellarini2015Non-localNon-Gaussianity}
\[ N_2(\bmath{k},\bmath{q}) = \frac{\bmath{k}\cdot \bmath{q}}{ k^2}. \]
Using the peak-background split, the two additional bias parameters, ($b_\Psi$, $b_{\Psi \delta}$), can be computed in Lagrangian space in terms of the density bias parameters and the matter variance at the scale $R$ of the tracers~\citep{Schmidt2010HaloNon-Gaussianity}
\[ \sigma_{R, \alpha}^2(k) = \frac{1}{2 \pi^2} \int dk_s \, k_s^2 \left(\frac{k}{k_s} \right)^\alpha P_R(k_s) \propto k^\alpha.  \label{eq:ScaleDepVar}\]
We choose for the smoothing scale $R$ of the power spectrum the Lagrangian radius corresponding to tracers of mass $M = 10^{13} M_\odot$. Using the Sheth-Tormen mass function and translating Lagrangian bias parameters to Eulerian space leads to the following expression for the first order bias parameter~\citep{Schmidt2010HaloNon-Gaussianity, Desjacques2011Non-gaussianThresholding, Schmidt2013Peak-backgroundClustering, Karagiannis2018ConstrainingSurveys}
\[ b_{\Psi}(k) = A f_{\text{NL}}^{\text{X}} \left[ 2 \delta_c (b_1-1) + 4 \left( \frac{d \ln \sigma_{R,\alpha}^2}{d \ln \sigma_{R,0}^2} -1 \right) \right] \frac{\sigma_{R,\alpha}^2}{\sigma_{R,0}^2}\,,\]
while the relevant second order parameter is given by~\citep{Giannantonio2010StructureSimulations, Karagiannis2018ConstrainingSurveys}
\begin{equation}
    \begin{split}
        b_{\Psi \delta}(k) = 2 A f_{\text{NL}}^{\text{X}} & \left[  \delta_c \left( b_2 + \frac{13}{21} (b_1 - 1) \right) \right. \\ & \left. + b_1 \left( 2 \frac{d \ln \sigma_{R,\alpha}^2}{d \ln \sigma_{R,0}^2} - 3 \right) + 1 \right] \frac{\sigma_{R,\alpha}^2}{\sigma_{R,0}^2} .
    \end{split}{}
\end{equation} 
The X serves as a placeholder for the local, equilateral and orthogonal templates. For local PNG, the scale dependent bias is caused by the Poisson factor needed to relate $\Psi$ with the (observed) density contrast $\delta$. For the equilateral and orthogonal PNGs this scale dependence is modified by the $\sigma_{R,\alpha}$ terms, see (\ref{eq:ScaleDepVar}), in the bias parameters.

The additional bias terms can be included into the $Z$ kernels which become~\citep{Tellarini2015Non-localNon-Gaussianity, Tellarini2016GalaxyDistortions}
\begin{subequations}
\begin{align}
    \begin{split}
        Z_1(\bmath{k}) &= b_1 + f \mu^2 +\frac{ b_\Psi(k) }{M(k)} \, ,
    \end{split} \\
    \begin{split}
        Z_2(\bmath{k_i}, \bmath{k_j}) = & b_1 F_2 + \frac{b_2}{2} + \frac{b_{s^2}}{2} S_2 + f \mu_{ij} G_2 \\ &+ f \, \frac{\mu_{ij} k_{ij}}{2} \left[ \frac{\mu_i}{k_i} Z_1(\bmath{k_j}) + \frac{\mu_j}{k_j} Z_1(\bmath{k_i}) \right] \\ 
        &+ \frac{1}{2} \left( \frac{(b_{\Psi \delta}(k_i)- b_{\Psi}(k_i) N(\bmath{k_j},\bmath{k_i}) )}{M(k_i)} \right. \\ 
        & \qquad + \left. \frac{(b_{\Psi \delta}(k_j) - b_{\Psi}(k_j) N(\bmath{k_i},\bmath{k_j}) )}{M(k_j)}\right).
    \end{split}
\end{align}
\end{subequations} 
For equilateral PNGs, the scale dependent bias becomes constant on large scales and thus degenerate with $b_1$. Moreover, its behavior on small scales is degenerate with derivative bias terms and probably unobservable~\citep{Assassi2015GalaxyNon-Gaussianity}. Hence, we exclude it from the power spectrum and bispectrum forecasts. 

We model the additional stochastic term, $\epsilon_\Psi$, as Poissonian shot noise which leads to the following non-vanishing power spectrum~\citep{Schmidt2016TowardsStructure, Desjacques2018Large-scaleBias}
\[ P_{\epsilon \epsilon_\Psi}(k) = \frac{b_\Psi(k) }{2 \bar{n}} .\]
All other correlators with the new stochastic term are either zero or higher order and thus discarded. This gives rise to three additional bias terms in the galaxy bispectrum
\[ B_{\text{ggg},\Psi -\text{SN} } =  \frac{b_1 b_\Psi(k_1) }{\bar{n} M(k_1)} P(k_1)+ 2 \, \text{perm}. \quad . \]

\subsection{Observed cross-spectra}
\label{ssec:obsSpectra}
Putting all the ingredients together, we can express the matter-galaxy cross-power spectra and bispectra. When projecting, the matter fields will correspond to the lensing convergence that does not suffer from RSDs, accordingly, we use the $F_2$ kernels. The galaxy fields in contrast, are affected by biasing, RSDs and scale-dependent bias from PNGs and described by the $Z_i$ kernels. The leading order power spectra are given by
\[  P_\text{gg}(k, \mu) = D_{\text{FoG}}^{\rm P}(k_{\parallel}) \left[  Z_1(k)^2 P(k) + \frac{1}{\bar{n}} \right] \quad, \]
\[P_\text{gm}(k, \mu) = \sqrt{D_{\text{FoG}}^{\rm P}(k_{\parallel})} Z_1(k)  P(k) \quad , \]
\[P_\text{mm}(k) =  P(k) \quad ,\]
where $\mu$ is the cosine of the angle of the wave vector with the LOS. The bispectra are given by
\[\begin{split}
B_{\text{ggg}} =  &D_{\text{FoG}}^{\text{B}}(k_{1,\parallel},k_{2,\parallel},k_{3,\parallel}) \times \\ &\biggl[ 2 Z_1(k_1) Z_1(k_1) Z_2(k_1, k_2) P(k_1) P(k_2) + 2 \, \text{perm.} \biggl. \\  \biggl. &+ Z_1(k_1) Z_1(k_2) Z_1(k_3)B_{\text{prim} }(k_1, k_2, k_3)  \biggr. \\  &+ \left. \frac{b_1}{\bar{n}} \left( Z_1(k_1) P(k_1)  + 2 \, \text{perm.} \right) + \frac{1}{\bar{n}^2}\right] ,\end{split}\]
\[ \begin{split}
B_{\text{ggm} }=  &D_{\text{FoG}}^{\text{B}}(k_{1,\parallel}, k_{2,\parallel} )  \biggl[2 Z_1(k_1) Z_2(k_1, k_3) P(k_1) P(k_3) \biggr. \\ 
&+ 2 Z_1(k_2) Z_2(k_2, k_3) P(k_2) P(k_3)  \\
&+ 2 Z_1(k_1) Z_1(k_2) F_2(k_1, k_2) P(k_1) P(k_2)  \\ 
& +  Z_1(k_1) Z_1(k_2)  B_{\text{prim}}(k_1, k_2,k_3)  \\ 
&+ \biggl. \frac{1}{\bar{n}}\left(b_1 +  \frac{b_\Psi(k_3)}{M(k_3)} \right) P(k_3) \biggr] , \end{split} \]
\[ \begin{split}
 B_{\text{gmm}} = &D_{\text{FoG}}^{\text{B}}(k_{1,\parallel}) \biggl[2 F_2(k_1, k_2) Z_1(k_1) P(k_1) P(k_2) \biggr.\\ 
 &+ 2 F_2 Z_1(k_1) P(k_1) P(k_3) \\ 
 &+ \biggl. 2 Z_2 P(k_2) P(k_3) + Z_1(k_1) B_{\text{prim}}(k_1, k_2,k_3) \biggr] , \end{split} \]
 \[ \begin{split}
 B_{\text{mmm}} =& \biggl[2 F_2 P(k_1) P(k_2)   + 2 \, \text{perm.}\biggr]+ B_{\text{prim} }(k_1, k_2,k_3). \end{split} \]
For the sake of compactness, we omitted the arguments of the bispectra. The matter bispectrum depends only on the magnitude of the three wave vectors. RSDs break the statistical isotropy and introduce an explicit dependence on the projection of the wave vector of the galaxy fields on the line of sight.
%%%%%%%%%%%%%%%%%%%%%%%%%%%%%%%%%%%%%%%%%%%%%%%%%%
%%%%%%%%%%%%%Projected Statistics %%%%%%%%%%%%%%%%
\begin{figure}
\centering
\includegraphics[width=\columnwidth]{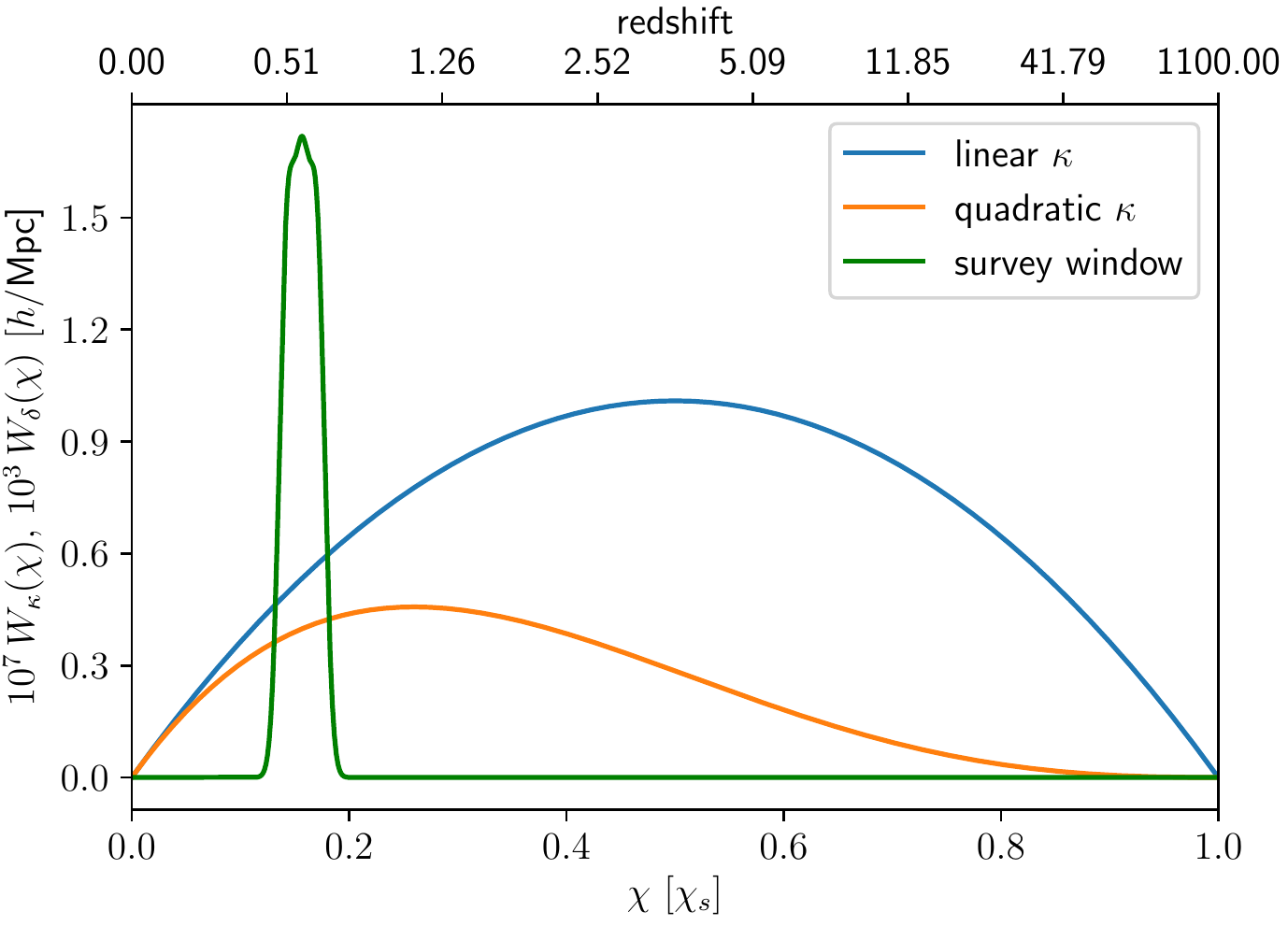}
\caption{Overview of the windows used in this paper. The CMB-lensing kernel (blue) peaks halfway between the observer and the surface of last scattering at $\chi_s$. The orange curve shows the lensing kernel multiplied with the growth factor. In green we show the sum of three galaxy bins with width and distance between the their centers of 200 \Mpc centered around z=0.57. The combined galaxy window shown here is constructed to to resemble the CMASS galaxy sample.}
\label{fig:LensingWindow}
\end{figure}

\begin{figure}
\centering
\includegraphics[width=\columnwidth]{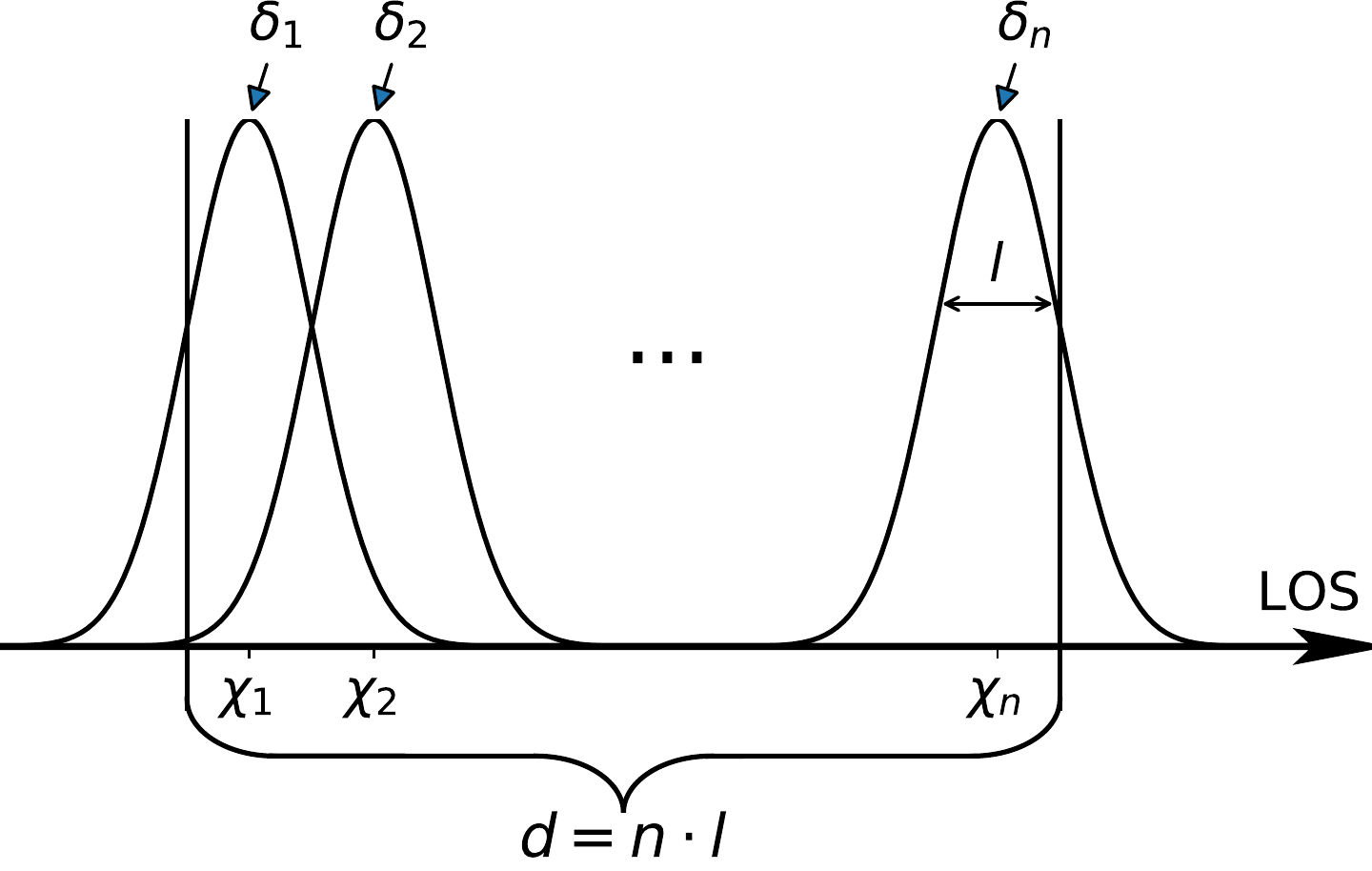}
\caption{We use a galaxy survey of fixed length $d$ consisting of $n$ identical tomographic bins along the line of sight. The center of each bin is given by~(\ref{eq:BinCenter}). Beside the Gaussian window that is shown here, we also use a top-hat window since it allows the survey volume to be covered evenly.}
\label{fig:SurveyBins} 
\end{figure}

\section{Projected statistics}

\label{sec:Projections}
In this section we derive the projection integrals for the power spectra and bispectra defined in subsection~\ref{ssec:obsSpectra}. Throughout this paper we use the flat sky approximation and project along the LOS. To simplify calculations, we neglect the time evolution within tomographic bins.

Fig.~\ref{fig:LensingWindow} gives an overview of the scenario we have in mind: The CMB lensing kernel determines how the matter between the surface of last scattering and the observer deflects photons from the CMB. In addition, we consider a galaxy survey at low redshift that allows to observe the late time universe directly and to cross-correlate CMB lensing and galaxy clustering. Fig.~\ref{fig:SurveyBins} shows the composition of galaxy survey window functions in more detail. We keep the survey's volume and position constant and vary the number of tomographic bins within the survey.

\subsection{Projections}
Projections of the matter density contrast along the LOS, $\delta$, with a window function $\tilde{W}$, are straightforward in real space:
\begin{equation}
    \begin{split}
 \delta_W(\bmath{x_{\perp}}) =& \int d x_{\parallel} \tilde{W}(x_{\parallel}) \delta(x_{\parallel}, \bmath{x_{\perp}})  \\ =& \int \frac{d^3 k}{(2 \pi)^3} d x_{\parallel} e^{-i \bmath{k \cdot x}}  \tilde{W}(x_{\parallel}) \delta(\bmath{k})  .
 \end{split}{}
 \end{equation}{}
This translates to Fourier space as 
\begin{equation}
\begin{split}
 \delta_W(\bmath{k_{\perp}}) =& \int \frac{d k_{\parallel}}{2 \pi} W(-k_{\parallel}) \delta(k_{\parallel}, \bmath{k_{\perp}})  \label{eq:ProjInt} .
\end{split}
\end{equation}
For galaxy clustering, we assume either Gaussian or Top-hat profiles whose functional forms read in Fourier space
\begin{subequations}
    \begin{eqnarray}
        W_{\text{TH}}^{\text{1D}}(k l; j) =& \text{sinc}(k l/2) \exp(i k \chi_j) \\
        W_{\text{G}}^{\text{1D}}(k l; j)  =&  \exp \left[-(k l/2)^2 /2 \right] \exp(i k \chi_j) . 
    \end{eqnarray}
\end{subequations} 
Here, the bin size, $l$, is related to the survey size, $d$, via $l = d/n$. The comoving position of the $j$-th bin center is given by
\begin{equation}
    \chi_j = \chi_c + \big[ j - (n+1)/2 \big] \,  l  \label{eq:BinCenter}
\end{equation}
where $\chi_c$ is the comoving distance to the center of the survey. The Top-hat window prevents volume effects when comparing 2D and 3D analysis or comparing 2D analysis with different numbers of bins (see Fig.~\ref{fig:SurveyBins}), but its slow decay for large wave vectors poses numerical challenges. The exponential decay of the Gaussian window in Fourier space in contrast allows us to project the galaxy bispectrum efficiently.

The lensing convergence
\begin{equation}
    \begin{split}
        \kappa(x_\perp) = & \frac{3}{2} \left(\frac{H_0}{c} \right)^2 \Omega_{m,0} \\
        & \int_{-\chi_s/2}^{\chi_s/2} d \chi \, \frac{(\chi +  \chi_s/2 ) ( \chi_s/2 - \chi)}{\chi_s} \frac{\delta(\chi+\chi_s/2)}{a},
    \end{split}
\end{equation}
describes the integrated effect of the matter fluctuations between the surface of last scattering surface at $\chi_s$ and the observer. In an Einstein-deSitter Universe, the scale factor, $a$, cancels the time evolution of the linear density contrast. The window function for the linear convergence field can then be expressed via a spherical Bessel function of the first kind
\[W_\kappa(k) = \frac{3}{2} \left(\frac{H_0}{c} \right)^2 \Omega_{m,0}\, \frac{\chi_s}{k} j_1(k \chi_s/2)  \exp\left[i k \chi_s/2 \right] \label{eq:kappa} .\]
For numerical reasons, we split the projection integral~(\ref{eq:ProjInt}) always into a time-dependent window function and a time-independent statistical field. This implies a different window function for each perturbative order of the convergence. In Appendix~\ref{sec:App1} we outline how to project higher orders of the lensing convergence in this framework.

\subsection{Projected power spectrum}
\label{ssec:PS}
The kernel of projected density fields is uniquely specified by $X_i \in (\delta,\kappa)$ and its positions $\chi_i$. Utilizing the symmetry of the window functions\footnote{This is correct for clustering and first order lensing.} makes cross-power spectra dependent on the two kernels involved and the distance between the centers of the two kernels. The projected power spectrum reads
\begin{equation}
    \begin{split}
        &P^{\text{2D}}_{X_1 X_2,|\chi_1-\chi_2|}(k_{\perp}) =  \langle \delta_{X_1,\chi_1}(\bmath{k_{\perp}}) \delta_{X_2,\chi_2}(\bmath{k_{\perp}'}) \rangle'  \\
        &=  2 \int_0^{\infty} \frac{d k_{\parallel}}{2 \pi} W_{X_1}(k_{\parallel})W_{X_2}(k_{\parallel})^* \cos \left[ |\chi_1-\chi_2| l k_{\parallel} \right] P\left(\sqrt{k_{\perp}^2+k_{\parallel}^2}\right).
    \end{split}
    \label{eq:PS2D}
\end{equation} 
 The apostrophe used after the correlator represents a suppressed factor of $(2\pi)^2 \delta^D(\bmath{k_{\perp}}+\bmath{k_{\perp}'})$, where $\delta^D$ is the Dirac delta. With $n$ tomographic bins, there are $n(n+1)/2$ different (cross-)galaxy power spectra, $n$ galaxy-CMB lensing spectra and one lensing-lensing power spectrum. 

\begin{figure}
\centering
\includegraphics[width=\columnwidth]{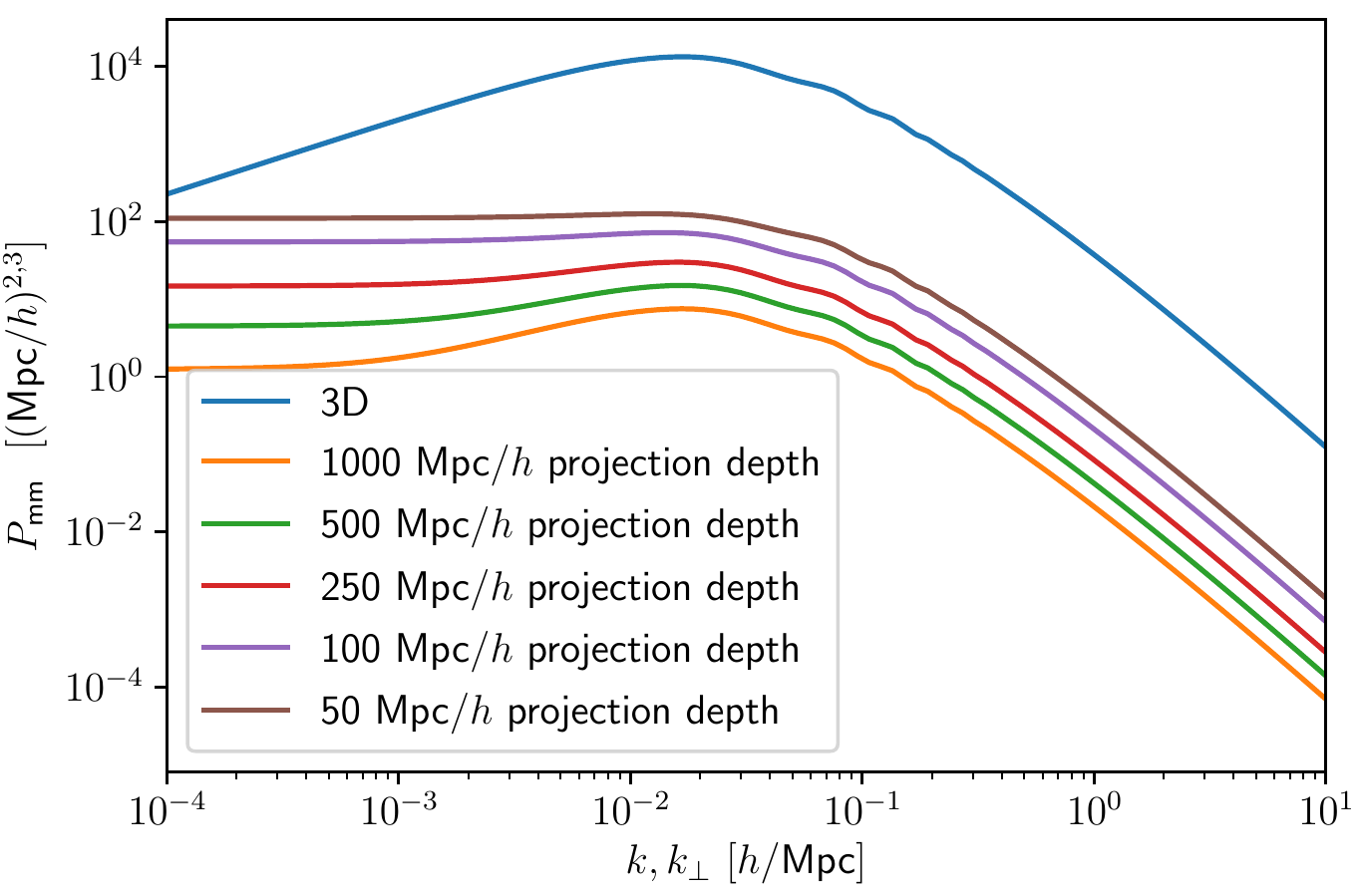}
\caption{Linear power spectrum (blue curve) and its 2D projections with a Gaussian kernel of varying depth (orange to brown). The projections change both the overall amplitude and the power spectrum's shape at low $k$. Instead of the $k^{n_s}$ scaling at low $k$, the projected power spectra become constant.}
\label{fig:Plin}
\end{figure}

The impact of the projection depth on the linear galaxy auto-power spectrum (i.e. $\chi_1=\chi_2$) is illustrated in Fig.~\ref{fig:Plin}. When comparing projected power spectra to the linear power spectrum in 3D, power spectra differ by an overall factor that comes from the volume of the window function in Fourier space. Since projections only source power from smaller to larger scales, the projections lead to an enhancement the large scales where the 3D power spectrum is increasing, i.e., for wavenumbers below the wavenumber corresponding to matter-radiation equality and the peak of the 3D power spectrum. Moreover, the strength of the effect is increasing with decreasing projection depth i.e. wider projection kernels. In the large wave number limit, the integrand of the projection integral (\ref{eq:PS2D}) becomes independent of the power spectrum and $P^{\rm 2D}(k)\propto P(k)$.

Since the projection of a 3D homogeneous, isotropic Gaussian random field yields a 2D homogeneous, isotropic Gaussian random field, the power spectrum's covariance can be computed as 
\begin{equation}
    \begin{split}
        \text{Cov}_{\text{G}}^{\text{2D}}[P^{\text{2D}}_{X_1 X_2|\chi_1-\chi_2|}&(k_{i,\perp}), P^{\text{2D}}_{X_3 X_4|\chi_3-\chi_4|}(k_{j,\perp})] = \\ 
        =\delta_{ij} \frac{k_f^2}{2\pi k_{i,\perp} \Delta k} & \left[ P^{\text{2D}}_{X_1 X_3 |\chi_1-\chi_3|}(k_{i,\perp}) P^{\text{2D}}_{X_2 X_4|\chi_2-\chi_4|}(k_{i,\perp}) \right. \\ 
        &\left. +P^{\text{2D}}_{X_2 X_3|\chi_2-\chi_3|}(k_{i,\perp}) P^{\text{2D}}_{X_1 X_4|\chi_1-\chi_4|}(k_{i,\perp})  \right]   .
    \end{split}  
    \label{eq:Cov-PS}
\end{equation}
The fundamental wavenumber of a quadratic survey with area $A$ is $k_\text{f} = 2 \pi/A^{1/2}$ and $\Delta k$ determines the bin size in $k$-space. The inclusion of cross-correlation between different tomographic bins is crucial to resolve all the information available in the survey.

\subsection{Projected bispectrum}
The projected cross-bispectrum correlating the density contrast of the tracers $\bmath{X} = (X_1, X_2, X_3) $ at positions $\bmath{\chi} = (\chi_1,\chi_2,\chi_3)$ is given by
\begin{equation}
\begin{split}
&B^{\text{2D}}_{\bmath{X},\bmath{\chi}}\left( k_{1, \perp},k_{2, \perp},k_{3, \perp} \right) \\ 
&= \langle \delta_{X_1 \chi_1}(k_{1, \perp})\delta_{X_2 \chi_2}(k_{2, \perp})\delta_{X_3 \chi_3}(k_{3, \perp}) \rangle' \\ 
=& \int_{-\infty}^\infty \frac{d k_{1, \parallel}}{2 \pi} \frac{d k_{2, \parallel}}{2 \pi} d k_{3, \parallel} \delta^D\left(\sum_i k_{i, \parallel} \right) \\
& W_{X_1 \chi_1}\left(k_{1, \parallel}\right) W_{X_2 \chi_2}\left(k_{2, \parallel}\right)W_{X_3 \chi_3}\left(k_{3, \parallel}\right)  \\  
& B\left( \sqrt{ k_{1, \perp}^2 + k_{1,\parallel}^2 },\sqrt{k_{2,\perp}^2 + k_{2, \parallel}^2 },\sqrt{ k_{3, \perp}^2 + k_{3, \parallel}^2 } \right) \\
=& \int_{-\infty}^{\infty} d x \int \prod_{i=1}^3 \left[\frac{d k_{i, \parallel}}{2 \pi} W_{X_i \chi_i}\left(k_{i, \parallel}\right) \exp \left(-i k_{i, \parallel} x \right) \right]  \times \\ \times & B\left( \sqrt{ k_{1, \perp}^2 + k_{1, \parallel}^2 },\sqrt{ k_{2, \perp}^2 + k_{2, \parallel}^2 },\sqrt{ k_{3, \perp}^2 +k_{3, \parallel}^2 } \right) .
\end{split}
\label{eq:BS2D} 
\end{equation}
From a numerical perspective, the last expression for the bispectrum is advantageous because it consists of three FFTs - whose results can be cached for each (window, wave vector) combination - followed by a 1D integration. In contrast, the full 2D integration is significantly slower and does not allow to use caching. From now on, we will drop both the subscripts $\perp$ and $\parallel$.
\begin{figure*}
\centering
\includegraphics[width=\textwidth]{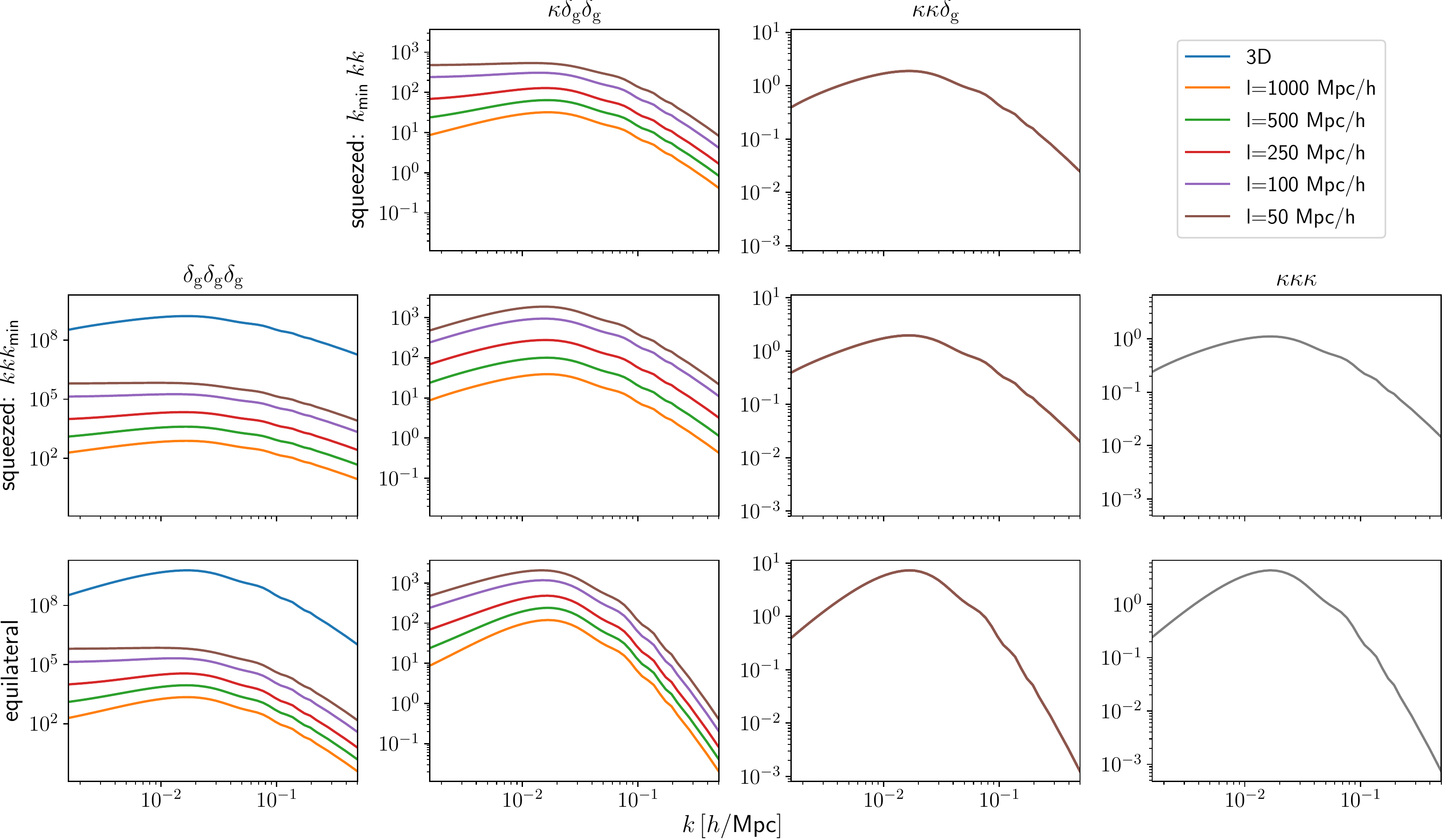}
\caption{The four columns show the four different (cross-)bispectra in the squeezed configuration (first two rows) and the equilateral configuration (bottom row). Where possible, we show the 3D bispectrum (blue) and projections with a Gaussian kernel of different depths (orange to brown). The lensing-lensing-clustering bispectra lie all on top of each other. }
\label{fig:crossBS}
\end{figure*}

Due to the angular dependence of the bispectrum, the 1D FFTs in the last part of equations~(\ref{eq:BS2D}) converge only for quickly decaying window functions. However, analytical progress can be made and we refer the interested reader to Appendix~\ref{sec:App1} for the details of the lensing projection integrals.

Fig.~\ref{fig:crossBS} compares the different projected bispectra (\ref{eq:BS2D}) for equilateral and squeezed configurations. We also show the 3D galaxy bispectrum and one sees that the projections have the strongest effect on largest scales. 

Following~\cite{Joachimi2009BispectrumLimit}, we compute the bispectrum's covariance in two dimensions as
\begin{equation}
\begin{split}
 \text{Cov}_G^{\rm 2D}&[B^{\rm 2D}_{\bmath{X},\bmath{\chi}}(k_1, k_2, k_3), B^{\rm 2D}_{\bmath{Y},\bmath{\psi}}(q_1, q_2, q_3)] =\\ &\delta^K_{\bmath{k, q}} \frac{(2 \pi) k_f^2}{k_1 k_2 k_3 (\Delta k)^3} \Lambda^{-1}(k_1, k_2, k_3) ~ \text{PPP}
\end{split}
\label{eq:Cov-BS}
\end{equation}
where 
\[ \Lambda^{-1}(k_1, k_2, k_3) = \frac{1}{4} \sqrt{2 k_1^2 k_2^2 + 2 k_1^2 k_3^2 + 2 k_2^2 k_3^2 - k_1^4 - k_2^4 -k_3^4}  \]
and
\begin{equation}
    \begin{split}
        \text{PPP} =& \sum_{(l,m,n)\in \sigma( \{1,2,3 \} )}\delta^K_{k_1, q_l} \delta^K_{k_2, q_m} \delta^K_{k_3, q_n} \times \\ \times&P^{\rm  2D}_{X_1 Y_l |\chi_1-\psi_l|}(k_1)P^{\rm  2D}_{X_2 Y_m |\chi_2-\psi_m|}(k_2)P^{\rm 2D}_{X_3 Y_n|\chi_3-\psi_n|}(k_3)   .
    \end{split}
\end{equation}
Here, $\delta^K$ refers to the Kronecker delta and the sum runs over all permutations of the bins $\psi_1, \psi_2, \psi_3$. For the galaxy bispectrum it is even more important than for the power spectrum to include cross-correlations between tomographic bins since neighboring bins are correlated and there are many cross-bin configurations that contribute.

\subsection{Theoretical uncertainties}
\begin{figure*}
\centering
\includegraphics[width=\textwidth]{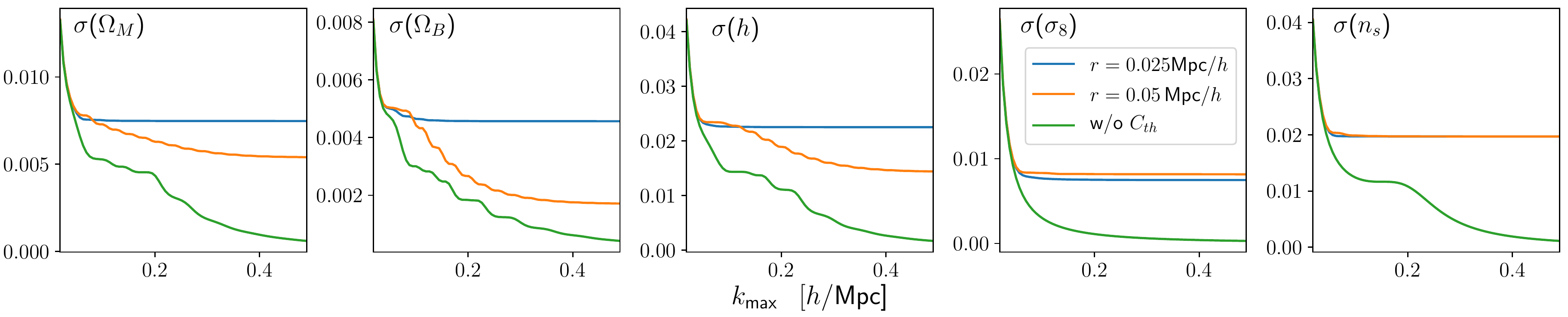}
\caption{Unmarginalized $1\sigma$ error bars without theoretical uncertainties (green line) and with theoretical uncertainties (orange and blue) as a function of the upper cut-off $k_{\text{max}}$. The error bars of the cosmological parameters are decreasing with $k_{\text{max}}$ without theoretical uncertainties. Using theoretical uncertainties, the error bars saturate. For a smaller correlation length in the correlation function, the saturation happens at a larger scale (blue curve) than with the larger correlation length (orange). The results come from a power spectrum analysis of a cubic survey of side length 1000\Mpc at redshift, z=0.57. }
\label{fig:GPcorrLen}
\end{figure*}

\begin{figure}
\centering
\includegraphics[width=\columnwidth]{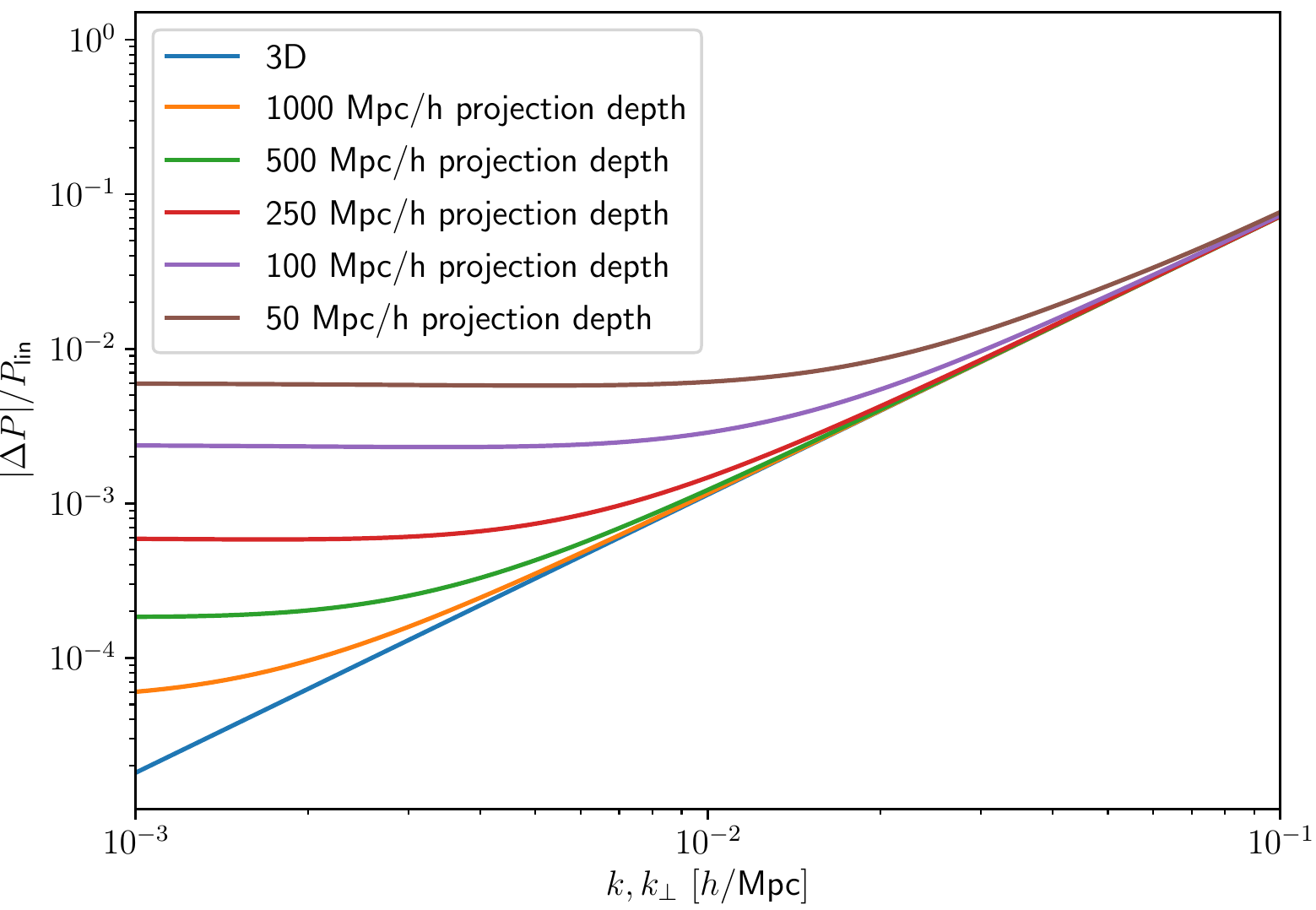}
\caption{Relative importance of the one-loop correction to the linear power spectrum in 3D (blue curve) and for various projected power spectra (orange to brown). The ratio in 3D follows a power law, whereas the ratios of the projected power spectra become constant at low $k$. At large $k$, the ratios of the projected power spectra approach the 3D value from above.}
\label{fig:ratio}
\end{figure}
\label{ssec:TheoUnc}
Typically, one deals with theoretical uncertainties due to the perturbative nature of the solutions of the evolution equations by means of a hard (possibly time dependent) cut-off, $k_{\text{max}}$, in the analysis. \cite{Baldauf2016LSSUncertainties} developed a more realistic approach based on the insight that the prediction's accuracy is lost gradually with increasing wave vectors and that the theoretical error can be estimated by the next-to-considered order of the perturbative expansion. We follow this proposal and model the theoretical uncertainties by a mean-zero Gaussian Process with Gaussian covariance function
\[ C_{\text{th}}(k_1, k_2) = E(k_1) \exp \left[ -\frac{(k_1-k_2)^2}{2 r^2} \right] E(k_2)   \]
of correlation length $r = r_{\text{bao}}/2$. We used half the correlation length than proposed by~\cite{Baldauf2016LSSUncertainties} which is motivated by the observation that the correlation length should be the length scale on which the spectra are roughly constant and not the one on which we observe changes. Since we work in this paper at leading order, we use the parametrisation for the envelope of the one-loop power spectrum from \cite{Baldauf2016LSSUncertainties},
\begin{equation} E(k) = b_1^2 \left( \frac{D(z)}{D(z_{\text{eff}} )} \right)^4 P(k) \left(\frac{k}{0.31 h \, \text{Mpc}^{-1} } \right)^{1.8} , \label{eq:PS1loopEnv}
\end{equation}
as the scale of uncertainty where $z_{\text{eff}}$ is the survey's mean redshift. Gaussian Processes have the handy property that marginalizing over them modifies the regular convariance function as follows
\begin{equation}
    C(k_1, k_2) \rightarrow C(k_1, k_2) + C_{\text{th} }(k_1, k_2) \quad .
\end{equation}
Fig.~\ref{fig:GPcorrLen} illustrates the effect of theoretical uncertainties on the forecasted errors (see section~\ref{ssec:FisherForecast}) in 3D. We compare the $k_{\text{max} }$ dependence of the marginalized error bars without theoretical uncertainties and with theoretical uncertainties of correlation lengths $r_{\text{BAO}}$ and $r_{\text{BAO}}/2$. In the latter case, the error bars saturate earlier than in the former case. 

Having outlined the formalism in 3D, we need to project the Gaussian Process. For the projection of the power spectrum covariance we refer to subsection~\ref{ssec:PS} and recall that the projection of a Gaussian Process remains a Gaussian Process. The key observation for the projection of the theoretical covariance, $C_{\text{th} }$, is that a Gaussian correlation function can be approximated by a finite sum
\begin{equation}
\begin{split}
&\exp \left[ -\frac{(k_1-k_2)^2}{2 r^2} \right] \\ &\simeq \frac{\sqrt{2}}{\sqrt{\pi} r }\Delta c \sum_{i=1}^{N} \exp \left[ \frac{ (k_1- (c_{\text{min}} +i \Delta c) )^2}{r^2} \right] \exp \left[ \frac{ (k_2-(c_{\text{min}}+i \Delta c) )^2}{r^2} \right],
\end{split}{}
\label{eq:CovThProj}
\end{equation}
where $c_{\text{min}}$ and $c_{\text{max}}$ are found empirically and $\Delta c = (c_{\text{max} }-c_{\text{min}})/N $. In practice, $\mathcal{O}$(100) terms are sufficient to achieve sub-percent precision. The separable representation has two advantages: instead of the 2D projection integral, one can perform 2N one dimensional integrations which is much faster in our setting. Moreover, the 1D integrals can be cached. This changes the scaling of the required integrations from quadratic to linear in $(k_{\text{max}}/\Delta k)$. Similarly, the scaling becomes linear in the number of redshift bins.

The above approach to the bispectrum's theoretical uncertainties becomes computationally intractable when projecting. Due to the implicit wavevector ordering in the correlation function, the direct projection is a 4 dimensional integral. As we were not able to speed the computations sufficiently up, we do not use theoretical uncertainties for the projected bispectrum.

To compare the effect of theoretical uncertainties in 2D and 3D, we compare the projected envelope of the one-loop power spectrum with the projected (linear) power spectrum to their 3D counterpart in Fig.~\ref{fig:ratio}. While the relative deviation of the one-loop envelope from the linear power spectrum follows a power law in 3D (see~(\ref{eq:PS1loopEnv})), the ratios in 2D deviate at small $k$ and become constant. For large $k$, the 2D ratios converge towards the 3D ratio from above.

%%%%%%%%%%%%%%%%%%%%%%%%%%%%%%%%%%%%%%%%%%%%%%%%%%
%%%%%%%%%%%%%%%%% Methodology %%%%%%%%%%%%%%%%%%%%
\section{Inference methods}
\label{sec:Metho}
In this section we first review Fisher forecasting and then outline our approach for computing parameter shifts due to inaccurate theoretical predictions.

\subsection{Fisher forecasting}
\label{ssec:FisherForecast}
The Cramer-Rao bound provides a lower bound on the statistical error for any unbiased, linear estimator in terms of the inverse of the Fisher Information (matrix)
\begin{equation}
 F_{ij} = \langle \left( \log \mathcal{L} \right)_{,ij} \rangle  ,
 \end{equation}
where $i,j$ label the parameters of interest, $\mathcal{L}$ is the likelihood function and all quantities are evaluated at the maximum-likelihood point. Assuming a Gaussian likelihood for the power spectrum and bispectrum, the Fisher Information can be calculated as~\citep{Tegmark1996Karhunen-LoeveSets}
\begin{equation}
    \begin{split}
        F_{ij} =& ~ \frac{1}{2}  \text{Tr} \left[ C^{-1} C_{,j} C^{-1} C_{,i} + C^{-1} (\mu_{,i} \mu_{,j}^T + \mu_{,j} \mu_{,i}^T )  \right] \\ \simeq& ~ \mu_{,i}^T C^{-1} \mu_{,j} \label{eq:FisherInfo}
        .
    \end{split}{}
\end{equation} 
The theory vector $\mu$ contains the spectra of interest and $C$ is the covariance. The derivatives in 3D are computed via finite differences~\citep{Smith2014PrecisionCodes}. Using the product rule and~\eqref{eq:PS2D}, \eqref{eq:BS2D} allows us to compute the derivatives in 2D. The (un)marginalized error forecasts are then given by
\begin{equation}
    \sigma^2_i = \begin{cases} 
      1/\text{F}_{ii} & \text{unmarginalized} \\
      \left(\text{F}^{-1}\right)_{ii} &  \text{marginalized} .
   \end{cases} \label{eq:ErrorBar}
\end{equation} 

\subsection{Parameter estimation}
\label{ssec:ParaEstimation}
We are interested in parameter shifts due to inaccurate (theoretical) modelling. In this scenario, we fit some theoretical model $\mu_{\bmath{\theta}}$ to the underlying ground truth $\mu_{\text{true}}$. Assuming a Gaussian distribution, this is done by choosing the parameters ${\bmath{\theta}}$ that maximize the following log-likelihood,
\[ -\chi^2 = -\frac{1}{2} ( \mu_{\text{true}} - \mu_{\bmath{\theta}} )^T C^{-1}  ( \mu_{\text{true}} - \mu_{\bmath{\theta}} ) + \text{const.} .\]
Since we investigate small biases, we can linearize the theoretical model around the best fit parameters ${\bmath{\theta}}_*$ as
\[ -\chi^2 = - \frac{1}{2} v^T C^{-1} v ,\]
where 
\[ v =  \left[ \mu_{\text{true}} - \mu_{{\bmath{\theta}}_*} - ({\bmath{\theta}}_* - {\bmath{\theta}})\cdot \left. \frac{\partial \mu_{\bmath{\theta}}}{\partial {\bmath{\theta}}} \right|_{{\bmath{\theta}}_*}  \right] . \]
In our case we use the ground truth parameters as best fit parameters. The likelihood of the linearized model has an explicit minimum
\[ {\bmath{\theta}} = {\bmath{\theta}}_* + F^{-1} \bmath{b} , \label{eq:Bias}\]
where $F$ is the Fisher Information and $\bmath{b}$ is given by
\[ \bmath{b} = \left( \mu_{\text{true}} - \mu_{{\bmath{\theta}}_*} \right) C^{-1} \left. \frac{\partial \mu_{\bmath{\theta}}}{\partial {\bmath{\theta}}} \right|_{{\bmath{\theta}}_*}  , \]
where for each component of b, one partial derivative is taken. (\ref{eq:Bias}) allows to compute the biases from using an inaccurate model. 

\subsection{Survey specifications}
For our forecasts and bias estimations, we assume a moderate-sized galaxy survey like CMASS-like with $n$ tomographic bins, each of depth $l=d/n$, where $d$ is the survey depth along the line of sight. Fig.~\ref{fig:SurveyBins} illustrates the setting and the details of the survey are given in Table~\ref{tab:kmax2Da}.

We use a $k$-space binning of $k_f$ (PS), $4 k_f$ (BS) and take three base points per fundamental frequency to obtain the averaged signal over $k$-intervals. We use the best-fit cosmology from~\cite{PlanckCollaboration2018PlanckParameters}: $\Omega_B = 0.0494, \Omega_M=0.3144 , h=0.6732$, \linebreak
$\sigma_8(z=0.57)=0.6029, n_s=0.966$.

%%%%%%%%%%%%%%%%%%%%%%%%%%%%%%%%%%%%%%%%%%%%%%%%%%
%%%%%%%%%%%%%%%%%%% Results %%%%%%%%%%%%%%%%%%%%%%
\section{Results}
\label{sec:Results}
In this section we first demonstrate empirically that one can recover the full 3D Fisher Information in projected surveys using a small enough projection depth. Next, we establish a consistent cut-off, $k_{\text{max}}^{\rm 2D}$, as a function of the number of redshift bins that allows to compare 2D surveys with different projection depths without the need of theoretical uncertainties. We then use this cut-off to analyse the error bar-bias trade off as a function of the FoG model~(\ref{eq:PS-FoG}, \ref{eq:BS-FoG}). We end the section with more optimistic forecasts.

Throughout this section, we use the finding that the cross-covariances between power spectra and bispectra are negligible at large scales~\citep{Song2015CosmologyBispectrum,Chan2017AnBispectrum, Yankelevich2019CosmologicalBispectrum}. When not stated otherwise, we use from section~\ref{ssec:CutOff} on a CMASS-like survey as described in Fig.~\ref{fig:LensingWindow} and Table~\ref{tab:kmax2Da}.

\subsection{3D-2D equivalence}
\begin{figure}
\centering
\includegraphics[width=\columnwidth]{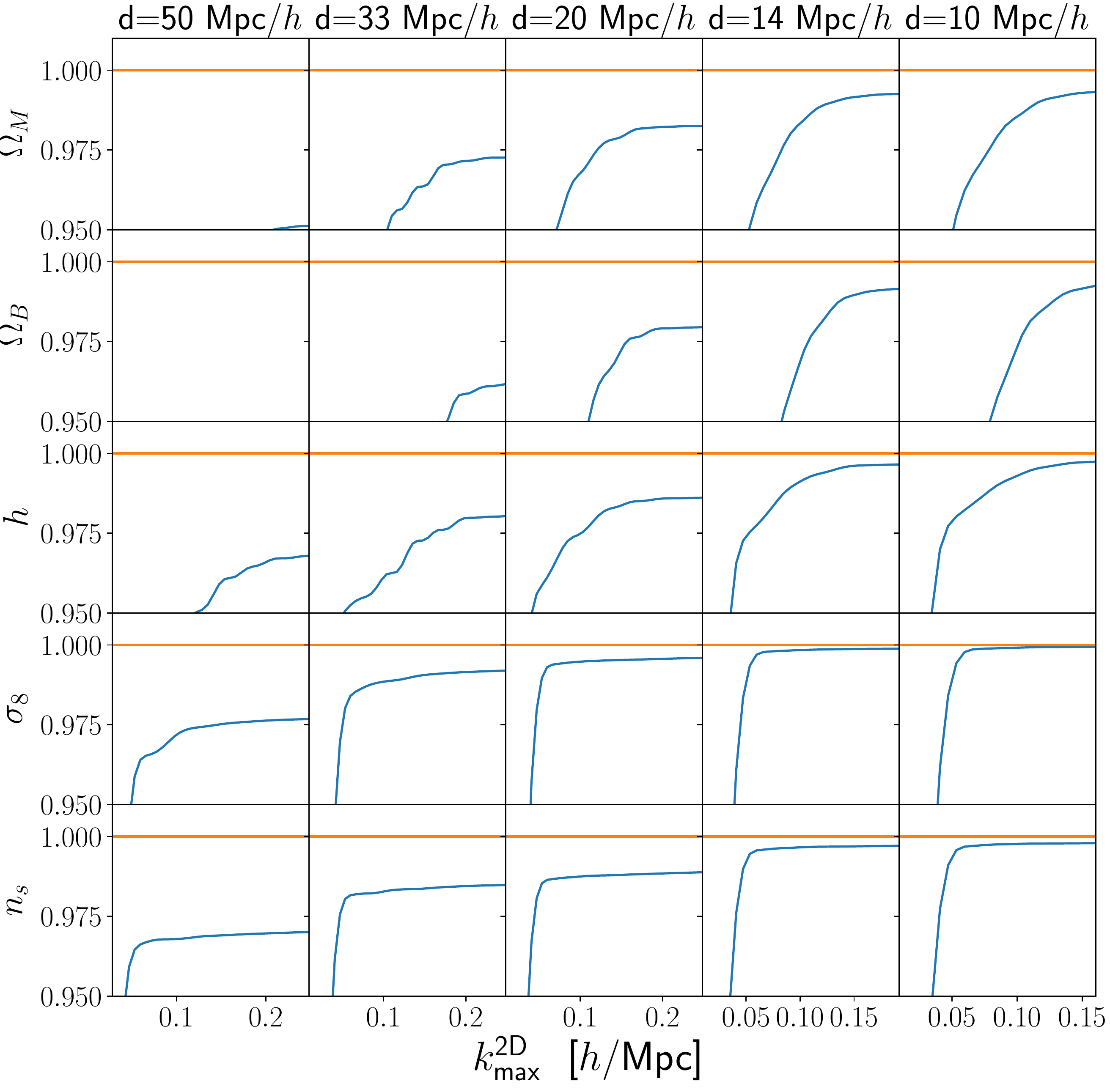}
\caption{The ratio of (unmarginalized) error bars in 3D and 2D, $\sigma_x^{\rm 3D}/\sigma_x^{\rm 2D}$, is shown as a function of $k_{\text{max}}^{\text{\rm 2D}}$. We use the saturated value for the 3D uncertainty and refer to Fig.~\ref{fig:GPcorrLen} where we studied the saturation in detail. For each parameter (row) and projection depth (column), the 2D-error bar saturates due to theoretical uncertainties. As the number of tomographic bins increases (from left to right), this saturated value approaches the 3D value. }
\label{fig:GP}
\end{figure}

We test the equivalence between three and two dimensional matter power spectrum analysis (without RSDs) empirically by performing both analyses and comparing the forecasted error bars. We control the sourcing of small-scale information to large scales with theoretical uncertainties. To this end, we use a cubic survey of side length $1000$ \Mpc. The corresponding fundamental frequency is roughly four times smaller than the correlation length of the theoretical uncertainties which ensures that the theoretical uncertainties are approximately constant over $k$-bin's with width of the fundamental frequency. The projections are done with Top-hat window functions of depths $1000/n$, where $n$ is the number of bins. Our findings are summarized in Fig.~\ref{fig:GP}. We show the ratio between the error bars in the two dimensional setting and the values the three dimensional analysis as a function of the 2D cut-off scale, $k_{\text{max}}^{\rm 2D}$. For a given number of bins, the error bars saturate in two dimensions, due to the projected theoretical uncertainties. In addition, those values converge to the 3D values as the number of bins increases.

\subsection{Choosing the cut-off scale for projected spectra}
\label{ssec:CutOff}
As explained in section~\ref{ssec:TheoUnc}, it is unfeasible to directly implement theoretical uncertainties for the projected bispectrum. Thus we need another approach to control for theoretical uncertainties in the matter predictions. In this work we control these systematics by choosing a cut-off scale, $k_{\text{max}}$, that ensures that all parameter shifts due to inaccurate matter modelling are below 20\% of the corresponding error bars. 

\begin{figure}
\centering
\includegraphics[width=\columnwidth]{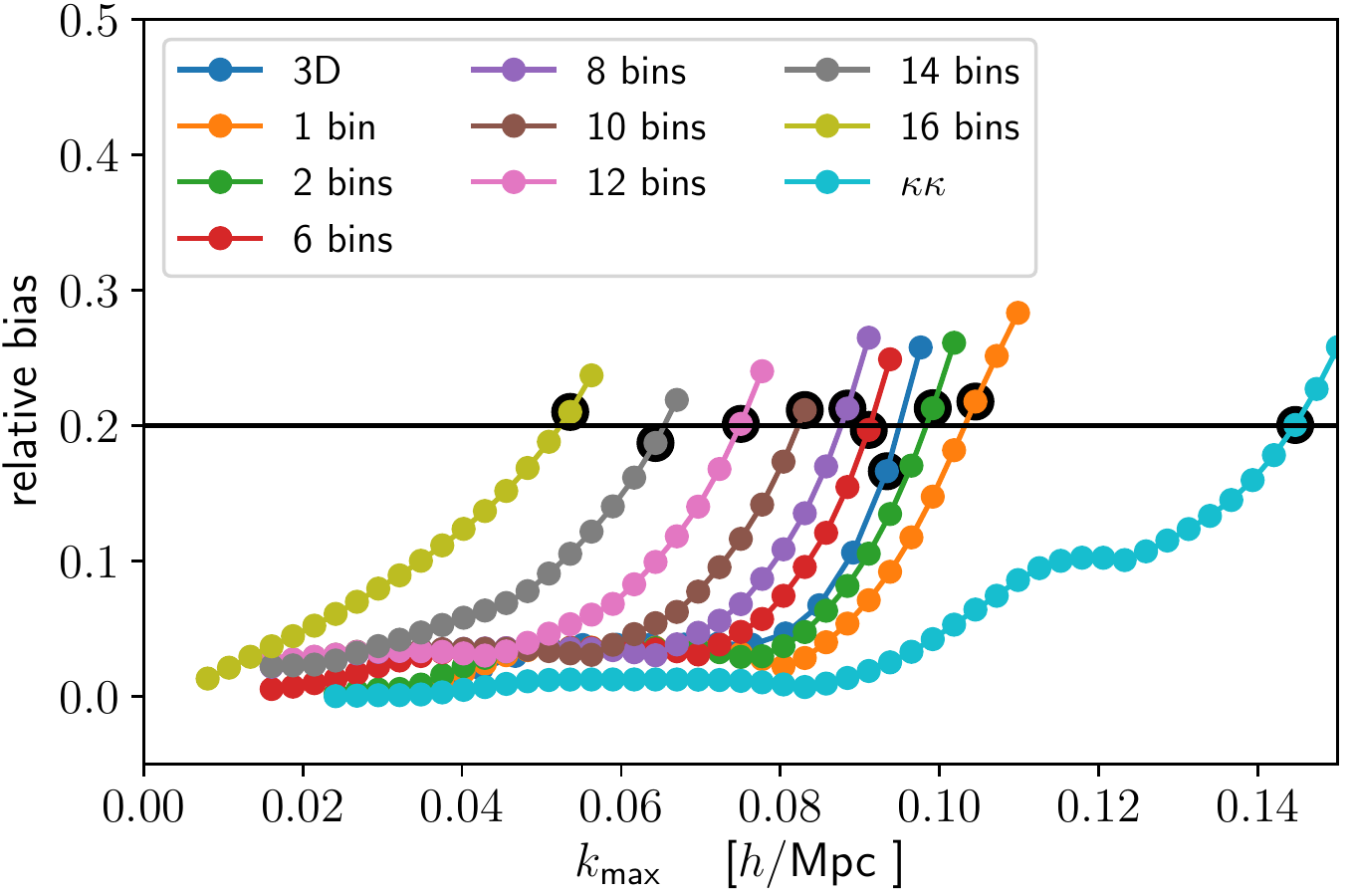}
\caption{Relation between $k_{\text{max}}$ and maximal relative bias for a varying number of redshift bins for a CMASS-like survey (see Table~\ref{tab:kmax2Da}). The points with the black border are closest to 0.2 for each configuration and thus used as cut-offs in this paper. Their numerical value can be found in Table~\ref{tab:kmax2Db}. The 3D curve was determined from a cubic survey of side length $l=(A\cdot d)^{1/3}$ with the same redshift and shot-noise as the 2D setting. }
\label{fig:kmax2D}
\end{figure}

We estimate the parameter shifts by fitting a linear matter power spectrum to the \textsc{halofit}~\citep{Smith2003StableSpectra} predictions as described in section~\ref{ssec:ParaEstimation}. Both models are without RSDs. Fig.~\ref{fig:kmax2D} illustrates the monotonic relation between cut-off scale and maximal relative biases for different projection depths. The cut-off is chosen to be the value where the maximal relative bias in the cosmological parameters is closest to 20\%. Those values are marked in black in the figure and the numerical values are reported in Table~\ref{tab:kmax2Db}. We want to point out, that the precise values of those cut-offs are specific for the chosen survey specified in Table~\ref{tab:kmax2Da}.

Since the amount of imprecise, small scale information that gets sourced to larger scales increases with decreasing projection depth, we see that the cut-off decreases as the number of tomographic bins increases. The effective cut-off in 3D, where no sourcing happens, lies in between those extremes because of the different $k$ dependence of the 2D and 3D covariance functions. The lensing cut-off is significantly larger for two reasons. Firstly, the kernel is very narrow in Fourier space and secondly, it peaks at early times, where non-linearities are small.

\begin{table}
\caption{(a): Characterisation of the CMASS-like survey we use in this paper. The velocity dispersion parameters were obtained from fits against $N$-body simulations. (b): cut-offs for different projection depths that ensure all relative biases in the matter predictions are below $20\%$ for a CMASS-like survey described on the right side. The 3D cut-off comes from a cubic survey with the same volume. The values correspond to the black points in Fig.~\ref{fig:kmax2D}. }
\subfloat[\label{tab:kmax2Da}]{
 \begin{tabular}{ l l }
    parameter & value  \\  [0.5ex] 
    \hline
    depth & 590 \Mpc  \\
    Area & 2345 $\text{Mpc}^2\, h^{-2}$   \\
     $z_\text{eff}$ & 0.57  \\
    $\bar{n}$ & $2\cdot 10^{-4}\, \text{Mpc}^3\, h^{-3}$ \\
    $b_1$ & $2.31$ \\
    $b_{s^2}$  &  $- \frac{4}{7}(b_1 - 1)$ \\
    $b_2$ & 0.77  \\
    $\sigma_P$ & 4 \Mpc \\
    $\sigma_B$ & $5.5$ \Mpc  \\
    \end{tabular}    }
    \subfloat[\label{tab:kmax2Db}]{
    \begin{tabular}{l c}
    Type & $k_\text{max}$ [\Mpc]  \\  [0.5ex] 
    \hline 
    3D$^*$ &  0.093 \\
    1 bin &  0.10 \\
    2 bins &  0.099 \\
    4 bins &  0.094 \\
    6 bins & 0.091   \\
    8 bins & 0.088  \\
    10 bins & 0.083  \\
    12 bins & 0.075  \\
    16 bins &  0.054 \\
    lensing & 0.14 \\
    \end{tabular} }
\end{table}

\begin{figure}
\includegraphics[width=\columnwidth]{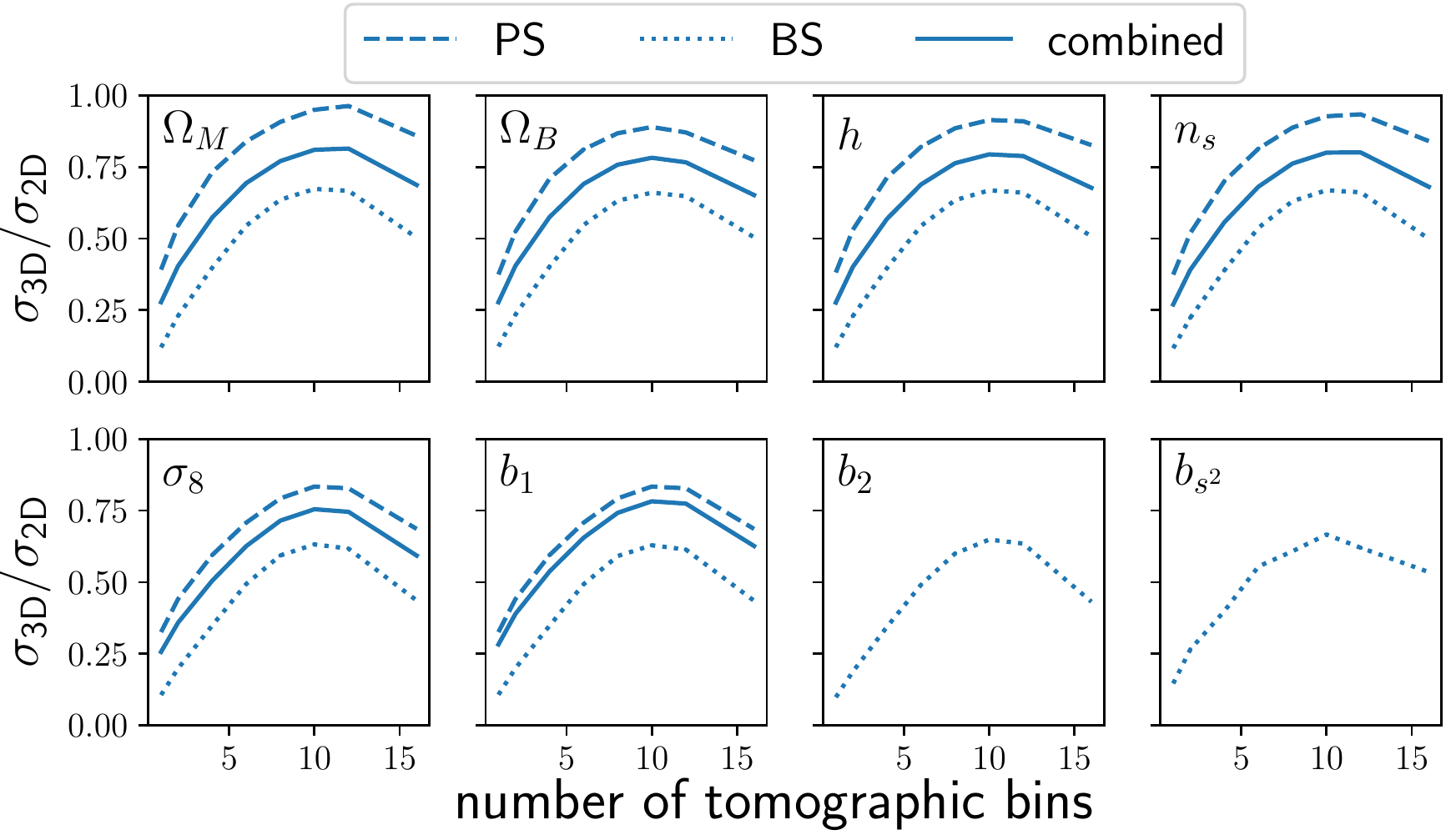}
\caption{The ratio of unmarginalized error bars in two and three dimensions is shown for the galaxy power spectrum (dashed), galaxy bispectrum (dotted) and combined (solid lines). With an increasing number of tomographic bins, the 2D error bars first decrease due to the increase in resolved radial information. Around 10 bins, the decrease in the cut-off $k_{\text{max}}^{\text{\rm 2D}}$ takes over and the error increases again.}
\label{fig:InfoLoss}
\end{figure}

In Fig.~\ref{fig:InfoLoss} we compare the forecasted error bars from the galaxy power spectrum and bispectrum using the chosen cut-offs in two and three dimensions. As the number of tomographic bins increases, one gains information by resolving more of the modes parallel to the line of sight from cross-correlations between the tomographic bins but looses at the same time from the overall decreasing cut-off scale. For both the power spectrum (dashed line) and the bispectrum (dotted) we see an increase in information until $\sim 10$ bins when the latter effects overtake and the information decreases again. This approach allows us to recover more than 80\% of bias/amplitude parameters and more than 90\% of cosmological parameters in a power spectrum analysis. In a pure bispectrum analysis, two thirds of the Fisher Information can be recovered compared to a 3D analysis. 

\begin{figure}
\centering
\includegraphics[width=\columnwidth]{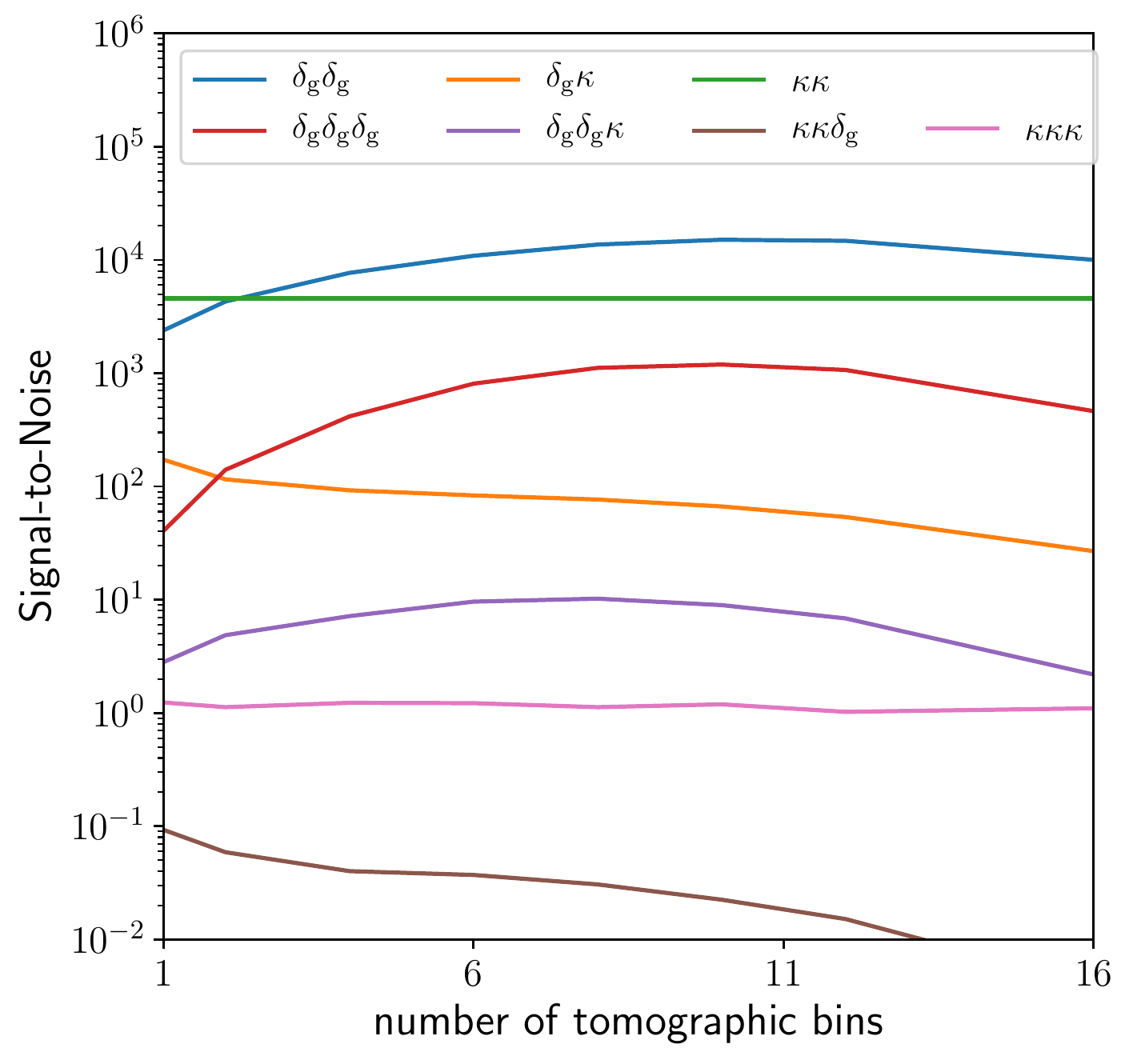}
\caption{Signal-to-noise scaling for different power specta and bispectra as a function of the number of tomographic bins. The dependence has three components: 1)  With more tomographic bins, more radial information is resolved. 2) The cut-off, $k_{\text{max}}^{\text{\rm 2D}}$, decreases with the number of tomographic bins which in turn decreases the SN. 3) The galaxy selection function is changing when using a variable number of  Gaussian profiles. This effect is negligible from four bins onward. }
\label{fig:SN}
\end{figure}
\subsection{Signal-to-noise}
\begin{figure*}
\centering
\includegraphics[width=\textwidth]{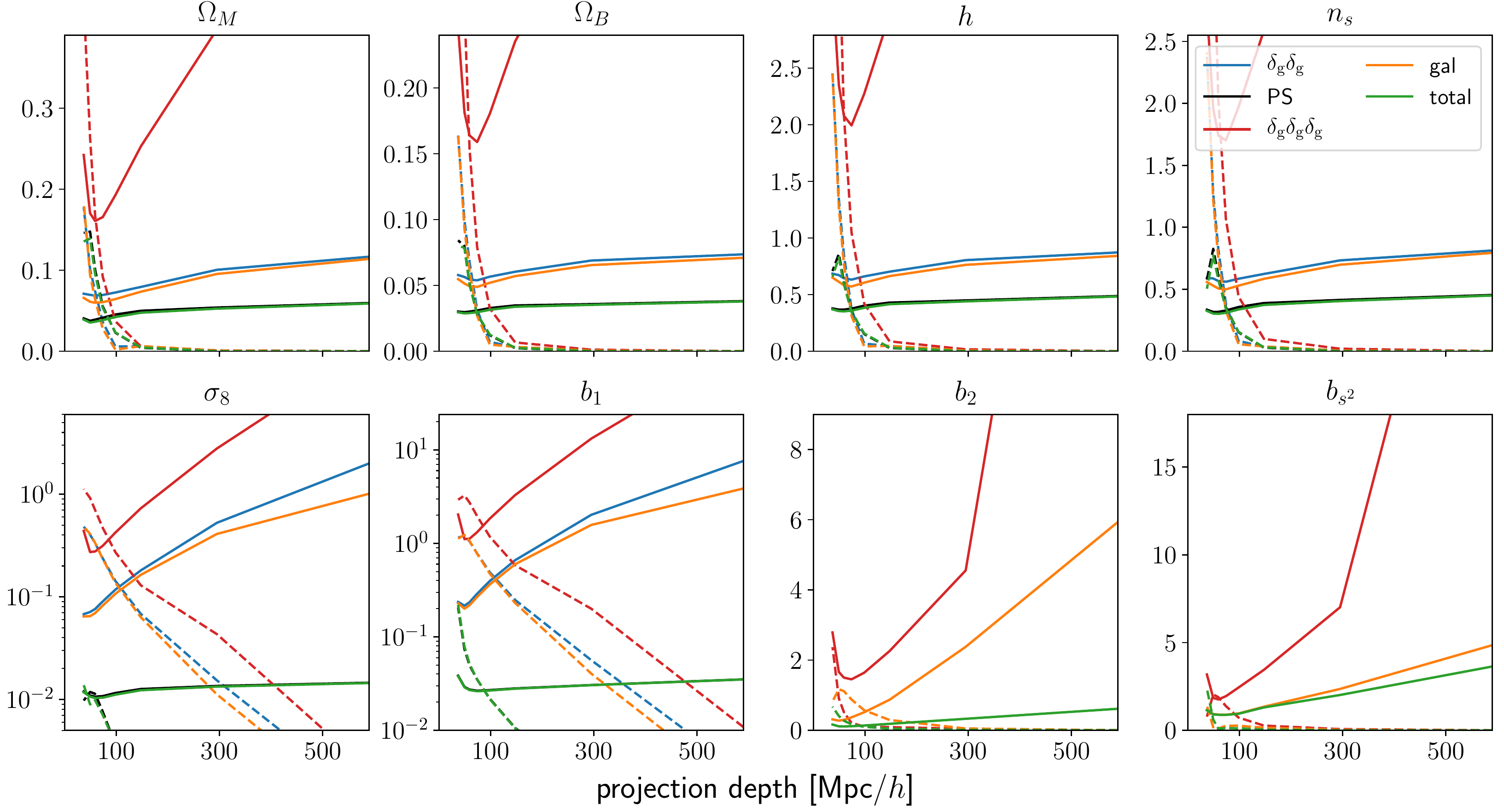}
\caption{Marginalized $1 \sigma$ error bars on cosmological parameters as a function of the projection depth for a selection of cross-spectra are displayed in solid lines. The Fisher forecasts assume FoG model 2, but are actually independent of the chosen FoG model. The corresponding dashed lines represent five times the bias due to inaccurate theoretical modelling for the FoG model that misspecifies the velocity dispersion by 50\%. All forecasted errors are minimal at $\sim 60$\Mpc. }
\label{fig:FisherInfo}
\end{figure*}
\begin{figure}
\centering
\includegraphics[width=\columnwidth]{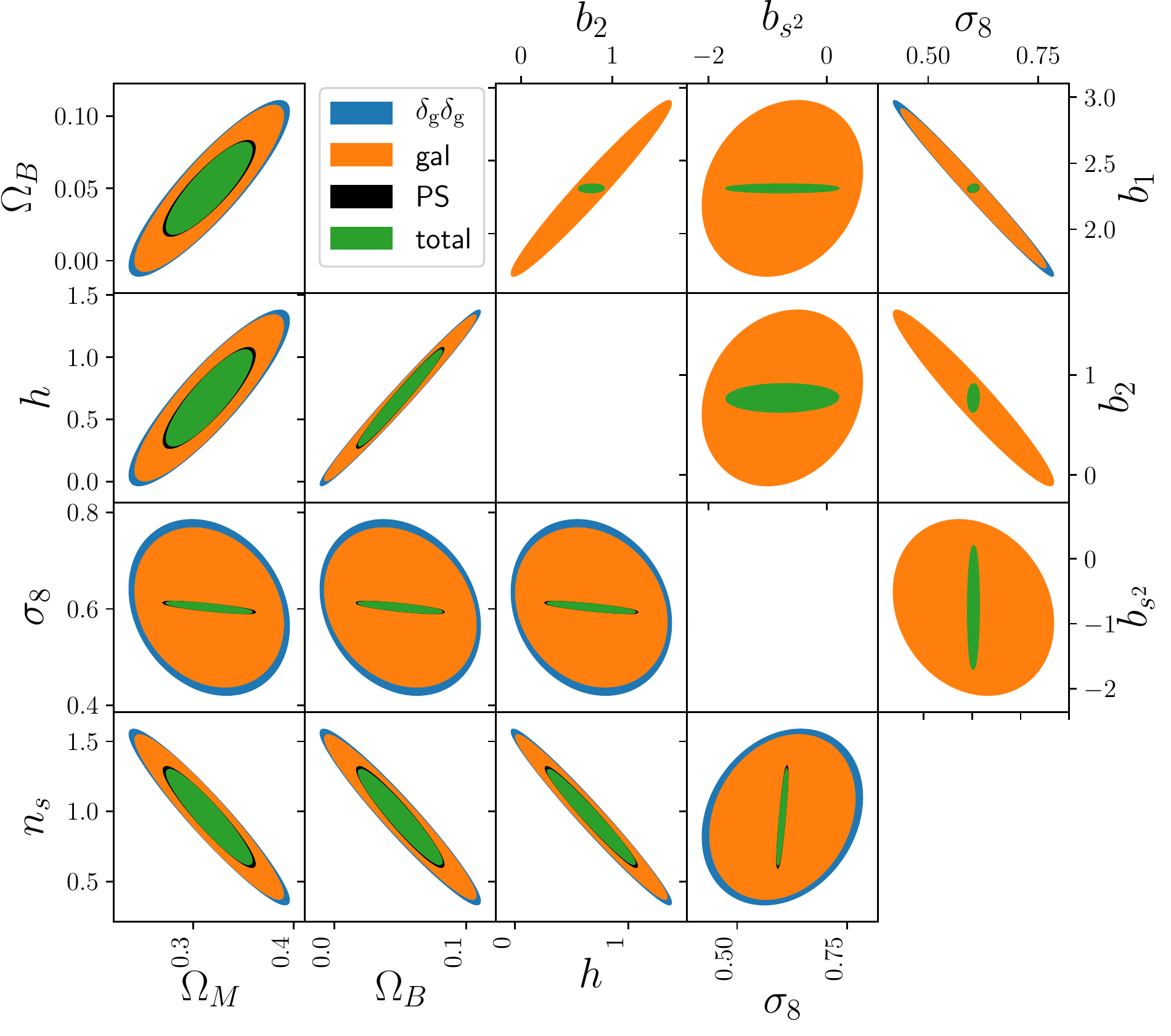}
\caption{The $1\sigma$ contour plots for four different spectra using FoG model 2. The bins are chosen to minimize the errors controlling for systematic uncertainties arising in FoG model 2 (velocity dispersion underestimated by 50\%). The optimal number of bins are reported in Table~\ref{tab:cosmoSum}.}
\label{fig:BestNBanana}
\end{figure}
There are three effects that determine the signal-to-noise (SN) scaling with respect to the number of tomographic bins. 1) As one increases the number of redshift bins, the projection depth decreases and more radial signal is resolved. 2) $k_{\text{max}}^{\rm 2D}$ decreases, if at least one galaxy field is involved, and with smaller projection depth, the SN decreases too. 3) The galaxy selection function is changing when using a variable number of  Gaussian profiles. Whereas the first two effects are relevant on all scales, the latter effect's size decreases with the number redshift bins, and is negligible from 4 bins onward. This justifies the use of Gaussian bins.

Fig.~\ref{fig:SN} displays the SN for all individual power specta and bispectra in our fiducial cosmology including RSDs. Due to our conservative cut-offs, the galaxy clustering and lensing auto-power spectra have the most SN. The galaxy bispectrum's SN is strongly increasing with the number of bins and nearly reaches the lensing power spectrum SN at its maximum at 10 bins. The cross-spectra tend to be much smaller than those auto-correlations since the overlap between the galaxy survey and the lensing kernel is small (see Fig.~\ref{fig:LensingWindow}). The lensing bispectrum SN shows small fluctuations since the binning of its (constant) cut-off is changing to ensure it is consistent with the changing galaxy clustering binning. 

Utilizing the covariance structure derived in section~\ref{ssec:PS}, there are analytical expressions for most of the power spectra's SN scalings and we report those in Appendix~\ref{sec:AppPS}. For the bispectrum, no analytical results exist. While for many power spectra the SN is independent of the chosen RSD model (see Appendix~\ref{sec:AppPS}) this property is approximately true for all the considered power specta and bispectra.

\subsection{Error - bias tradeoff}
In this subsection we investigate the error-bias trade-off in two scenarios. First we assume a $\Lambda$CDM cosmology and forecast the parameter uncertainties and shifts due to inaccurate FoG modelling. Next, we perform a similar analysis for the bias and PNG parameters for a fixed cosmology. 

In both cases, we assume a ground-truth Gaussian FoG model with velocity dispersions specified in Table~\ref{tab:kmax2Da}. We then study inaccurate models by changing the velocity dispersion parameters by $-100\% $ (i.e. no FoG modelling), $-50\%, -10\%, 10\%$ and $50\%$. Since we find that the biases are approximately independent of the sign of the shift in the velocity dispersions, we restrict ourselves to the first three cases and refer to them as model 1-3. The statistical uncertainties that we report together with biases are always the ones obtained from the inaccurate FoG model. In practice however, the uncertainties are nearly independent of the chosen FoG model.

\begin{table*}
    \caption{This table summarises the optimal forecasted relative errors for three FoG models considered. For each spectra we give the number of tomographic bins such that the maximal relative bias is below 20\% and the statistical uncertainties are minimal. On the left side, we report the marginalized relative $1\sigma$ uncertainties. On the right side, we show the corresponding relative biases.}
    \label{tab:cosmoSum}
    \begin{tabular}{ l c c c c c c c c c c c}
    \hline
    & \multicolumn{5}{c}{relative $1\sigma$ uncertainties   [\%]} &  & \multicolumn{5}{c}{relative biases  [\%]}  \\ [0.5ex] 
     & $\delta_{\text{g}} \delta_{\text{g}}$ &PS & $\delta_{\text{g}} \delta_{\text{g}} \delta_{\text{g}}$ & gal& total & \hspace{0.7cm}  & $ \delta_{\text{g}} \delta_{\text{g}}$ & PS & $\delta_{\text{g}} \delta_{\text{g}} \delta_{\text{g}}$ & gal & total \\ [0.5ex] 
    \hline
    \multicolumn{12}{c}{Model 1: $\sigma_v = 0 \cdot \sigma_{ \text{v, ground truth}}$}  \\ [0.5ex]
    \hline
    bins     & 4 & 4 & 6 & 4 & 4 &&&&& \\
$\Omega_M$ &  25 & 16 & 60 & 23 & 15 & & 1.9 & 2.2 & -5.4 & 2.3 & 2.6 \\
$\Omega_B$ & 120 & 70 & 360 & 110 & 68 & & 1.4 & 1.8 & -5 & 1.5 & 2.0 \\
$h$        & 100 & 64 & 330 & 99 & 62 & & 1.7 & 1.7 & -5.2 & 1.9 & 2.0 \\
$n_s$      & 65 & 40 & 200 & 61 & 39 & &-1.5 & -1.9 & 6.2 & -1.7 & -2.2 \\
$\sigma_8$ & 30 & 2.1 & 65 & 27 & 2.0 & &10 & -1.4 & 18 & 10 & -1.6 \\
$b_1$      & 29 & 1.2 & 74 & 26 & 1.2 & & -10 & 10 & -18 & -10 & 10 \\
$b_2$      &   &   & 210 & 110 & 24 &  &  &   & -1.9 & -9.1 & 6.7 \\
$b_{s^2}$  &   &   & 320 & 180 & 170 &  & &   & 8.1 & -2.9 & -0.8 \\
   \hline 
   \multicolumn{12}{c}{Model 2: $\sigma_v = 0.5  \cdot \sigma_{ \text{v, ground truth}}$} \\ [0.5ex] 
    \hline
    bins & 4 & 6 & 6 & 4 & 6 &&&&&&   \\
   $\Omega_M$& 25 & 14 & 61 & 23 & 13 && 1.5 & 10 & -3.8 & 1.7 & 11 \\
   $\Omega_B$& 120 & 66 & 360 & 110 & 63 & & 1.0 & 7.4 & -3.6 & 1.1 & 8.0 \\
   $h$       &    100 & 60 & 340 & 99 & 57 & & 1.3 & 7.5 & -3.7 & 1.4 & 8.2 \\
   $n_s$     &    65 & 37 & 210 & 61 & 35 &&  -1.1 & -8.5 & 4.5 & -1.3 & -9.2 \\
   $\sigma_8$& 30 & 1.9 & 70 & 27 & 1.8 && 7.5 & -6.2 & 13 & 7.7 & -6.5 \\
   $b_1$     &    29 & 1.2 & 80 & 26 & 1.2 && -7.5 & 16 & -13 & -7.7 & 16 \\
   $b_2$     &      &   & 210 & 110 & 18 &  & &   & -1.3 & -6.8 & 15 \\
   $b_{s^2}$ &   &   & 330 & 180 & 130 &   & &  & 5.9 & -2.2 & -2.4 \\
   \hline
   \multicolumn{12}{c}{Model 3: $\sigma_v = 0.9 \cdot \sigma_{ \text{v, ground truth}} $ } \\ [0.5ex] 
    \hline
    bins & 8 & 10 & $10^*$ & 8 & 10 &&&&&&   \\
   $\Omega_M$& 22 & 12 & 50 & 19 & 11 & & -2.6 & 13 & -5.1 & -2.5 & 13 \\
   $\Omega_B$& 110 & 60 & 330 & 97 & 58 & & -3 & 8.7 & -4.6 & -2.9 & 9.0 \\
   $h$       &    93 & 54 & 310 & 85 & 52 & & -3 & 8.5 & -4.9 & -2.9 & 8.6 \\
   $n_s$     &    58 & 33 & 180 & 52 & 31 & & 3.5 & -9.8 & 5.7 & 3.4 & -9.9 \\
   $\sigma_8$& 15 & 1.8 & 41 & 14 & 1.7 & & 13 & -5.4 & 14 & 15 & -4.9 \\
   $b_1$     &    12 & 1.2 & 44 & 11 & 1.2 & & -14 & 9.3 & -13 & -15 & 9.6 \\
   $b_2$     &      &   & 200 & 48 & 14 &   & &  & -2.1 & -12 & 12 \\
   $b_{s^2}$ &   &   & 240 & 120 & 120 &   &  & & 5.0 & -1.4 & 0.5 \\
   \hline
    \end{tabular}   
\end{table*}

Moreover, we only show the results of the following five spectra combinations that we consider most interesting: galaxy power spectrum ($ \delta_{\text{g}} \delta_{\text{g}}$), all power spectra combined (PS), galaxy bispectrum ($\delta_{\text{g}} \delta_{\text{g}} \delta_{\text{g}}$), galaxy power spectrum and bispectrum combined (gal) and all power spectra plus bispectra combined (total).
\begin{figure*}
\centering
\includegraphics[width=\textwidth]{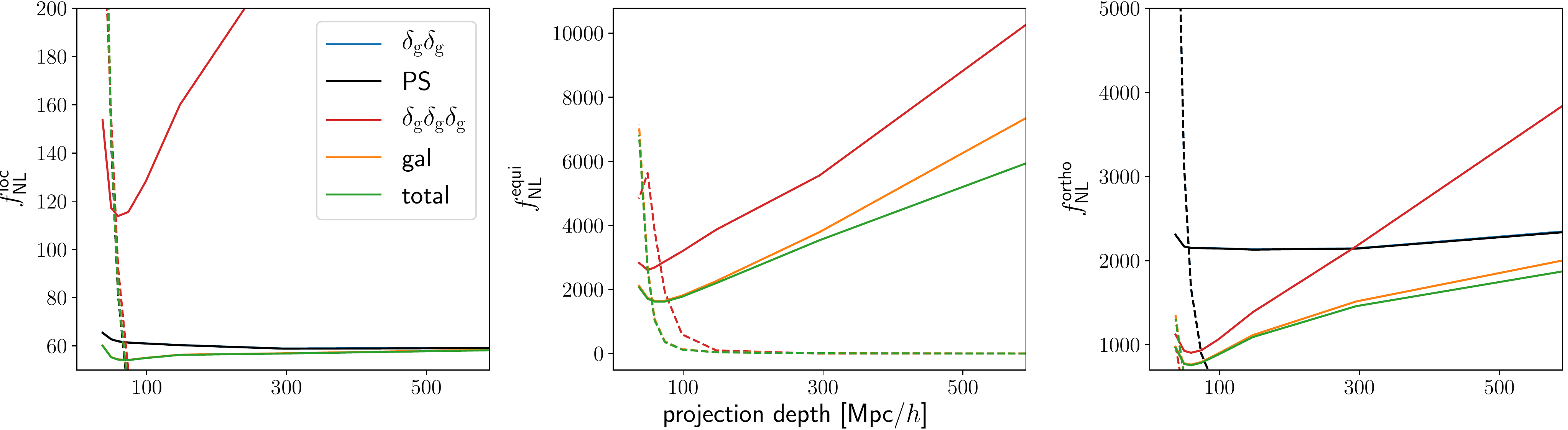}
\caption{The marginalized $1\sigma$ error bars on PNGs as a function of the projection depth for a selection of cross-spectra are displayed in solid lines. The Fisher forecasts assume FoG model 2, but are almost independent of the chosen FoG model. The corresponding dashed lines represent five times the bias due to inaccurate theoretical modelling for the FoG model that misspecifies the velocity dispersion by 50\%. Each subplot corresponds to a different forecasts and we marginalized over the bias parameters. }
\label{fig:fnlscaling}
\end{figure*}
\subsubsection{Cosmological forecasts}
\label{ssec:tradeoff}
 We fit the three above mentioned FoG models to the fiducial model and report the error bars and relative biases for all parameters. Fig.~\ref{fig:FisherInfo} shows the scaling of the relative biases and error bars for the model where the velocity dispersion is $50\%$ smaller than its fiducial value (model 2). The five combinations can roughly be grouped into three sets. The galaxy bispectrum has the least constraining power of the considered spectra and the derived error bars decrease very strongly with the number of tomographic bins until they reach a minimum around 10 bins (projection depth $l\simeq 60$ \Mpc). For smaller projection depths, the error bars increase again due to the decreasing $k_{\text{max} }^{\text{2D}}$. The galaxy power spectrum and the combination of galaxy power spectrum and bispectrum scales more weakly with the number of resolved bins and have their minimum around 10 bins too. Finally, adding information from galaxy lensing to either the galaxy power spectrum alone or both power spectrum and bispectrum allows us to lower error bars once more. This is partly because CMB-lensing allows us to break the $b_1$-$\sigma_8$ degeneracy. It does not help to constrain $b_{s^2}$ better, as the trace-free part of the tidal tensor is rather uncorrelated to the trace of the tidal tensor that corresponds to the amplitude/bias parameters, which CMB-lensing can constrain well. The relative biases remain close to zero for all parameters, with the exception of $\sigma_8$ and $b_1$, until a projection depth of $l\sim100$\Mpc, when they start growing quickly. Changing the FoG model, has a marginal impact on the error bars but shifts the curves of the relative biases left (right) if the FoG model becomes more (less) accurate.

In Table~\ref{tab:cosmoSum} we summarize our findings for the optimal error bars (left side) together with the relative biases (right side) conditioned on being smaller than $20\%$. We observe that the error bar difference across the spectra is significantly larger than the difference within the spectra across FoG models. 

The decrease in Fisher Information for projection depths $ \leq60$\Mpc implies that even with a velocity dispersion that is off by $10\%$, one is able to fully recover the available information in two dimensions. In contrast, a misspecification of $50\%$ leads to a $\sim10\%$ increase in the error for cosmological parameters and more for bias parameters. The worst case scenario of not modelling the FoG effect at all, inflates the error bars by $\sim20\%$. 

The relative biases tend to be strongest in parameters that affect the amplitude strongly and tend to be positive, since the models considered underestimate the FoG damping.

The 2D contour plots of the best case Fisher Information matrix of model 2 (see Table~\ref{tab:cosmoSum}) is displayed in Fig.~\ref{fig:BestNBanana}. One sees that all spectra have approximately the same covariance structure and that CMB lensing helps to break the $\sigma_8$-$b_1$ degeneracy. The positive correlation between $b_1$ and $b_2$ is explained as follows: The three bias/amplitude parameters $\sigma_8$, $b_1$ and $b_2$ are all pairwise anticorrelated. However, the positive definiteness of the covariance matrix pushes the weakest among them, $b_2$-$b_1$, to a positive value in the joint analysis.

\subsubsection{PNG forecasts}
Assuming a known cosmology, we now forecast bias and $f_{\text{NL}}$ parameters for the local, equilateral and orthogonal shape. We perform a separate forecast for each template and FoG model and illustrate the dependence on the number of tomographic bins in Fig.~\ref{fig:fnlscaling}. 

The scale dependent bias in the galaxy power spectrum yields the best constraints for the local shape. Since the survey is roughly four times as wide as deep, the constraints from the power spectrum have no dependency on the projection depth. The error from the bispectrum analysis, in contrast, shows a strong dependence on the projection depths. For few bins, the error is very large but decreases quickly with decreasing projection depths. As with the cosmological parameters, the error bars become worse for more than 10 bins due to the decreasing cut-off. Since the equilateral shape does not lead to scale dependent bias, it can only be constrained from the bispectrum and the power spectra only contribute towards reducing the uncertainty in the bias parameters. The constraints for the orthogonal shape from the power spectrum and bispectrum are of similar order, since the scale dependent bias only scales as $1/k$. 

The relative biases are close to zero for large projection depths and only start playing a role around 100 \Mpc. We report the optimal forecasts with relative biases below $20\%$ in Table~\ref{tab:PNGSum}. We observe that the error bar differences across spectra are larger than across FoG models given a spectra. For the local shape, the latter differences are basically zero, since only the largest scales are relevant. The constraints of the equilateral and orthogonal shape in contrast improve by $\sim10\%$ when modelling the FoG effect precisely.
\begin{figure}
\centering
\includegraphics[width=\columnwidth]{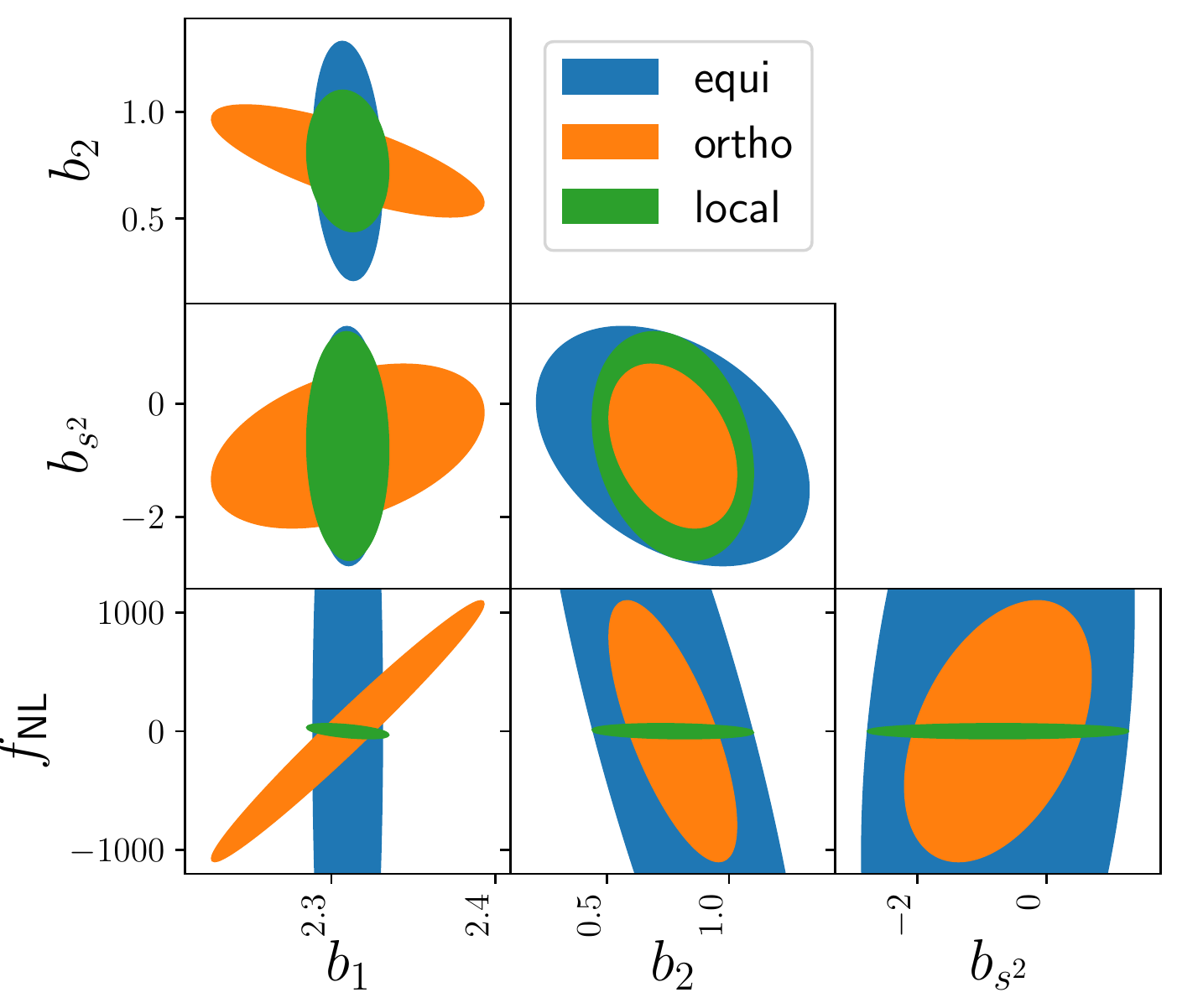}
\caption{$1\sigma$ contour plots for an $f_{\text{NL}}$+bias forecast combining all power spectra and bispectra and using FoG model 2. The bins are chosen to minimize the errors controlling for systematic uncertainties that arise in FoG model 2 (velocity dispersion underestimated by 50\%). Each color corresponds to an independent forecast.}
\label{fig:fnlmarginal}
\end{figure}

\begin{table*}
    \caption{Optimal forecasted errors for a bias-$f_{\rm NL}$ analysis with fixed cosmological parameters. We present separate forecasts for the three FoG models considered. For each spectra we give the number of tomographic bins such that the maximal relative bias is below 20\% and the statistical uncertainties are minimal. On the left side, we report the marginalized $1\sigma$ uncertainties. On the right side, we show the corresponding relative biases.}
    \label{tab:PNGSum}
    \begin{tabular}{ l  c c c c cccc c c c}
    \hline
    & \multicolumn{5}{c}{$1\sigma$ uncertainties} & & \multicolumn{5}{c}{relative biases   [\%]}  \\ [0.5ex] 
     & $ \delta_{\text{g}} \delta_{\text{g}}$ & PS & $\delta_{\text{g}} \delta_{\text{g}} \delta_{\text{g}}$ & gal  & total & \hspace{0.3cm} & $ \delta_{\text{g}} \delta_{\text{g}}$ & PS & $\delta_{\text{g}} \delta_{\text{g}} \delta_{\text{g}}$ & gal & total \\ [0.5ex] 
    \hline  
    \hline 
    \multicolumn{12}{c}{Model 1: $\sigma_v = 0 \cdot \sigma_{ \text{v, ground truth}}$} \\ [0.2ex] 
    \hline 
    bins & 2 & 2 & 6 & 2 & 2 & & & & & & \\ 
    $f_{\text{NL}}^{\text{local} }$ & 60 & 60 & 120 & 56 & 56 && -1.4 & -1.4 &-1.1 &-1.7 & -1.7\\
    $b_1$                   & 0.020 &  0.020 & 0.29 &  0.020 &  0.020 &&  16 &  16 &- 0.47 & 17 & 17 \\
    $b_2$                   & & & 1.1 & 0.18 & 0.18 &&  & & 11 & 7.2 & 7.2\\
    $b_{s^2}$               & & & 1.3 & 1.3 & 1.3   &&  & & 0.81 &-0.79 & -0.79\\
    \hline
    bins &  &  & 6 & 2 & 2 & & & & & & \\ 
    $f_{\text{NL}}^{\text{equi} }$ & & & 2900 & 2300 & 2200 && & &-18 & -0.51 &-0.47 \\
    $b_1$                  & & & 0.51 & 0.016 & 0.016 &&  &  &-15 & 20 & 20 \\
    $b_2$                  & & & 2.2 & 0.36 & 0.35 && & & 20 & 3.9 &  4.0 \\
    $b_{s^2}$              & & & 1.6 & 1.4 & 1.4 && & &  9.1 &-0.92 & -0.90 \\
    \hline
    bins & 6 & 6 & 10 & 4 & 4 & & & & & & \\ 
    $f_{\text{NL}}^{\text{ ortho} }$ & 2200 & 2200 & 920 & 900 & 880 &&  11 &  11 &  12 & 4.5 & 4.4 \\
    $b_1$                   & 0.16 &  0.16 &  0.42 &  0.068 &  0.067 &&  19 &  19 & 17 & 15 & 15 \\
    $b_2$                   & & & 1.4 & 0.21 &  0.21 &&   &  &-4.4 & 8.9 & 9.0 \\
    $b_{s^2}$               & & & 1.4 & 1.1 & 1.1 && &  & 3.7 &-0.88 &-0.95 \\
    \hline  
    \hline 
    \multicolumn{12}{c}{Model 2: $\sigma_v = 0.5 \cdot \sigma_{ \text{v, ground truth}}$} \\ [0.2ex]
    \hline 
    bins & 2 & 2 & 8 & 2 & 2 & & & & & & \\ 
    $f_{\text{NL}}^{\text{local} }$ & 60 & 60 & 110 & 56 & 56 &&-1.1& -1.1 &-1.6 &-1.3 &-1.3 \\
    $b_1$                   & 0.020 & 0.020 & 0.30 & 0.020 &  0.020 &&  12 &  12 & 0.064 & 12 & 12 \\
    $b_2$                   & & & 1.1 & 0.18 & 0.18 &&  & &  11 & 5.4 & 5.4 \\
    $b_{s^2}$               & & & 1.3 & 1.3 & 1.3 && & & 4.1 & -0.60 &-0.59 \\
    \hline
    bins &  &  & 6 & 2 & 2 & & & & & & \\ 
    $f_{\text{NL}}^{\text{equi} }$ & & & 2900 & 2300 & 2200 && & & -13 & -0.38 & -0.35 \\
    $b_1$                  & & & 0.51 & 0.016 & 0.016 &&  &  & -11 &  15 &  15 \\
    $b_2$                  & & & 2.2 & 0.36 & 0.35 && & & 15 & 2.9 & 3.0 \\
    $b_{s^2}$              & & & 1.6 & 1.4 & 1.4  &&  &  & 6.8 & -0.69 & -0.67\\
    \hline
    bins & 6 & 6 & 10 & 4 & 4 & & & & & & \\ 
    $f_{\text{NL}}^{\text{ortho} }$ & 2200 & 2200 & 930 & 900 & 890 && 8.3 &  8.3 & 8.6 & 3.3 & 3.3 \\
    $b_1$                   & 0.16 & 0.16 & 0.42 & 0.068 & 0.067 && 15 &  15 & 13 & 11 & 11 \\
    $b_2$                   & & & 1.4& 0.21 & 0.21 && & & -3.1 & 6.6 & 6.7 \\
    $b_{s^2}$               & & & 1.4& 1.1 & 1.1 && & & 2.8 &-0.66 &-0.71 \\
    \hline 
    \hline 
    \multicolumn{12}{c}{Model 3: $\sigma_v = 0.9 \cdot \sigma_{ \text{v, ground truth}}$} \\ [0.2ex] 
    \hline 
    bins & 6 & 6 & 10 & 6 & 6 & & & & & & \\ 
    $f_{\text{NL}}^{\text{local} }$ & 61 & 61 & 120 & 54 & 54 && -3.2 & -3.2 &-4.1 & -3.8 &-3.8 \\
    $b_1$                   & 0.017 & 0.017 & 0.31 & 0.016 & 0.016 &&  17 &  17 &-0.035& 18 & 18\\
    $b_2$                   & & & 1.1 & 0.12 & 0.12 && & & 2.9 &  8.0 & 8.0 \\
    $b_{s^2}$               & & & 1.3 & 0.88 & 0.87 && & & 1.1 & -1.1 &-1.1 \\
    \hline
    bins &  &  & 10 & 4 & 4 & & & & & & \\ 
    $f_{\text{NL}}^{\text{equi} }$ & & & 2700 & 1800 & 1800 &&  &  & -7.3& -0.38 &-0.36 \\
    $b_1$                  & & & 0.50 & 0.013 & 0.013 && & &-5.9 &  10 & 10\\
    $b_2$                  & & & 2.1 & 0.28 & 0.28 && & & 7.5&  2.1 & 2.1 \\
    $b_{s^2}$              & & & 1.5 & 1.0 & 1.0 && & & 3.9& -0.76 &-0.75 \\
    \hline
    bins & 10 & 10 & 10 & 10 & 10 & & & & & & \\ 
    $f_{\text{NL}}^{\text{ortho} }$ & 2200 & 2200 & 910 & 790 & 780 &&  7.4 &  7.4 & 1.1&  4.2 & 4.2 \\
    $b_1$                   & 0.16 & 0.16 & 0.37 & 0.059 & 0.058 && 11 & 11 & 0.52 & 14 & 14 \\
    $b_2$                   & & & 1.3 &  0.19 &  0.18 && &  & 1.4 & 4.3 & 4.4\\
    $b_{s^2}$               & & & 1.3 & 1.2 & 1.2 && & & 0.69 & 3.5 & 3.5 \\
    \hline 
    \end{tabular}   
\end{table*}
\begin{table*}
    \caption{Forecasts for the relative $1\sigma$ uncertainties of cosmological parameters in a CMASS-like survey with $k_{\text{max}}=0.15$ \Mpc and 6 tomographic bins without (left side) and with (right) CMB prior. }
    \label{tab:Cosmokmax}
    \begin{tabular}{ l c c c c c c c c c c c}
    \hline
    & \multicolumn{5}{c}{relative $1\sigma$ uncertainties} &  & \multicolumn{5}{c}{relative  $1\sigma$ uncertainties}  \\
    & \multicolumn{5}{c}{without CMB prior   [\%]} &  & \multicolumn{5}{c}{with CMB prior   [\%]}  \\ [0.5ex] 
     & $ \delta_{\text{g}} \delta_{\text{g}}$ &PS & $\delta_{\text{g}} \delta_{\text{g}} \delta_{\text{g}}$ & gal& total & \hspace{0.7cm}  & $ \delta_{\text{g}} \delta_{\text{g}}$ & PS & $\delta_{\text{g}} \delta_{\text{g}} \delta_{\text{g}}$ & gal & total \\ [0.5ex] 
    \hline
   $\Omega_M$& 13 & 11 & 22 & 10 & 9.1 &&2.1 & 1.9 & 2.5 & 1.9 & 1.8\\
   $\Omega_B$& 50 & 43 & 91 & 41 & 37 &&1.1 & 1.0 & 1.3 & 1.0 &  0.95\\
   $h$       & 47 & 40 & 85 & 38 & 34  &&0.76 & 0.70 & 0.90 & 0.70 & 0.66\\
   $n_s$     & 31 & 26 & 57 & 24 & 22 &&0.42 & 0.40 & 0.46 & 0.41 & 0.39\\
   $\sigma_8$& 15 & 1.3 & 31 & 11 & 1.2 &&0.88 & 0.57 & 0.88 & 0.88 & 0.57 \\
   $b_1$     & 15 & 0.8 & 36 & 11 & 0.8 && 0.96 & 0.68 & 4.7 & 0.96 & 0.67\\
   $b_2$     &   &   & 80 & 51 & 10 &&  &   & 63 & 10 & 9.9 \\
   $b_{s^2}$ &   &   & 110 & 56 & 51 &&  &   & 75 & 50 & 50  \\
   \hline
    \end{tabular}   
\end{table*}

\begin{table}
    \caption{Forecasts for the $1\sigma$ uncertainties of $f_{\text{NL}}$ and bias parameters in a CMASS-like survey with $k_{\text{max}}=0.15$ \Mpc and 6 tomographic bins.}
    \label{tab:PNGkmax}
    \begin{tabular}{lc c c c c }
    \hline 
     & $ \delta_{\text{g}} \delta_{\text{g}}$ &PS & $\delta_{\text{g}} \delta_{\text{g}} \delta_{\text{g}}$ & gal& total  \\ [0.5ex] 
     \hline
    $f_{\text{NL}}^{\text{local} }$ & 57 & 57 & 76 & 45 & 44 \\ 
    $b_1$           & 0.012 & 0.012 & 0.11 &  0.011 & 0.011\\ 
    $b_2$           &  & & 0.50 & 0.075 & 0.075\\ 
    $b_{s^2}$       &  & & 0.56 & 0.38 & 0.38\\ 
    \hline 
    $f_{\text{NL}}^{\text{equi} }$ &  &  & 2300 & 1200 & 1200\\ 
    $b_1$                  &  &  & 0.20 & 0.0088 & 0.0088 \\
    $b_2$                  &  &  & 1.1 &  0.13 & 0.13 \\
    $b_{s^2}$              &  &  & 0.68 & 0.40 & 0.40 \\ 
    \hline
    $f_{\text{NL}}^{\text{ortho} }$ & 1750 & 1750 & 680 & 630 & 620 \\ 
    $b_1$                   & 0.13 & 0.13 & 0.12 & 0.049 & 0.048 \\ 
    $b_2$                   & & & 0.52 & 0.15 & 0.15  \\ 
    $b_{s^2}$               & & & 0.56 & 0.39 & 0.39  \\ 
    \hline 
    \end{tabular}
\end{table}{}
\subsection{Optimistic forecast}
\label{ssec:OptimisticForecast}
We believe an optimistic but still realistic scenario is given by an upper cut-off of $k_{\text{max}}^{\text{2D} }=0.15$\iMpc and 6 redshift bins. This is in line with the choice of~\cite{Karagiannis2018ConstrainingSurveys, Yankelevich2019CosmologicalBispectrum}. \cite{Agarwal2020} were able to work with larger cut-offs for the power spectrum by using separate values for the power spectrum and bispectrum and computing parameter shifts (\ref{eq:Bias}) relative to the next perturbative order instead of the fully non-linear \textsc{halofit} prediction. In Table~\ref{tab:Cosmokmax} we summarize the error bars for this scenario in our fiducial cosmology with RSDs. Since, the cut-off for CMB lensing has previously been close to 0.15\iMpc, there are no significant improvements there. In contrast, the error bars from the galaxy power spectrum shrink by a factor of 2 and for the bispectrum by a factor of 3 compared to the best case forecasts with the conservative cut-off from Table~\ref{tab:kmax2Db}. The combined error bars from an analysis with all spectra shrink by $30\%$. Adding a CMB prior leads to dramatic improvement for all parameters that can be constrained from CMB observations. This is expected because only the next generation galaxy surveys will contain a comparable information content to current CMB surveys. The CMB prior is based on Appendix A of~\cite{Smith2014PrecisionCodes}. Let us stress that CMASS is a galaxy sample from a moderate-sized survey volume and thus constraints from this sample should not be expected to be competitive with Planck constraints. Upcoming surveys will map regions that are 50-100 times larger than CMASS. This is expected to translate into seven to ten times tighter parameter constraints. 

Table~\ref{tab:PNGkmax} contains our PNG forecasts in the optimistic scenario. Since PNGs are best determined on large scales, adding small scale information decreases the constraints only by $\sim20\%$. Since the large volume of future surveys translates into a smaller $k_\text{f}$, constraints on local and orthogonal type non-Gaussinity are expected to improve by more than the volume related factor of seven to ten mentioned above. This is due to the scale dependent bias, which dominates on large scales and leads to additional survey volume dependence.

\subsection{A simple signal compression approach}
The computations can be sped up by removing the configurations from the analysis that contain the least information. Since most configurations contributing to the Fisher matrix are cross-correlations between distinct tomographic galaxy bins, we use the following simple implementation of the idea. Given a correlator with two or three galaxy fields, we only include configurations where the maximal distance between the two galaxy bins is less than some threshold. This reduces the scaling of bispectrum configurations from cubic to linear in the number of tomographic bins. Optimizing the threshold, we find that this speeds up computations, with and without RSDs, for 16 bins by a factor of more than 10 while still recovering more than 99\% of the Fisher Information in all parameters. For a smaller total number of bins (deeper bins), the sped-ups are smaller. However, it is not possible to improve the bias-error trade-off for any of the considered FoG models. For further details we refer to Appendix~\ref{App:MCL}.
%%%%%%%%%%%%%%%%%%%%%%%%%%%%%%%%%%%%%%%%%%%%%%%%%%
%%%%%%%%%%%%%%%%%% Conclusion %%%%%%%%%%%%%%%%%%%%

\section{Conclusion}
\label{sec:Conc}
In this paper we quantified the statistical power of two- and three-point correlators of projected density fields in constraining cosmological parameters and primordial non-Gaussianity. We investigated the trade-off between statistical errors and biases induced by imperfect modelling of non-linear redshift space distortions, in particular the Finger-of-God effect. We developed an efficient implementation of the projection integrals required to predict projected power spectra and bispectra.

Using a model for theoretical uncertainties, we have shown empirically that one can recover the full 3D Fisher Information in tomographic surveys with sufficiently small projection depths. Using a projection depth of $l=10$\Mpc allows us to recover 99\% of the 3D information. The full account for theoretical uncertainties is numerically not feasible for the 2D bispectrum due to the large number of four dimensional projection integrals that are needed for the non-sparse covariance matrix of the theoretical uncertainties. Instead, we control theoretical uncertainties by computing cut-offs that depend on the projection depth. Those cut-offs are chosen such that the maximal relative biases due to inaccuracies in the matter predictions are less than 20\%. This approach allows to recover more than 80\% of the information in bias/amplitude parameters and more than 90\% of the information in cosmological parameters in a power spectrum analysis. In a bispectrum analysis, this approach allows us to recover 70\% of the 3D Fisher Information.

Next, we studied the relation between FoG modelling and relative biases in the forecasted parameters for a CMASS-like survey. We found that the resulting biases are independent of whether one over- or underestimates the FoG damping. We found that not modelling the FoG effect inflates error bars by $20\%$ when controlling for biases. A model that underestimates the velocity dispersion by $50\%$ leads to an increase of $10\%$ and a model whose velocity dispersion differs by $10\%$ is able to recover the full information while maintaining all relative biases smaller than 20\%. 

We performed a similar analysis for PNGs of the local, equilateral and orthogonal shape. Here, the necessity to model the FoG effect depends crucially on whether or not one best constrains the PNGs from the scale dependent bias or from the template. Whereas in the former case, one can recover most of the information without modelling non-linear RSDs, one can improve the error bars by 10\% and 20\% for the template dominated orthogonal and equilateral shape respectively with an accurate FoG model.

In a more optimistic scenario where one can control systematic uncertainties up to $k_{\text{max}}=0.15$\Mpc, one can expect further improvements of more than $100$\% for spectra that contain galaxy clustering information. This translates into $30\%$ improvements of combined (PS, total) error bars. Let us stress that future surveys would further improve constraints by 700-1000\% due to the much larger volumes enabling LSS constraints as tight as those provided by Planck. With these surveys, the projected power spectrum and bispectrum provide a conservative yet powerful analysis toolkit.

Lastly, we studied the impact of dropping cross-bin galaxy correlations. We find that this is not a tool to improve the bias-error trade-off. However, it is possible to reduce the number of clustering bispectrum configurations considerably without losing much Fisher Information. In practice we were able to speed up the most numerically demanding configurations by a factor of more than 10 while losing less than 1\% of Fisher Information in all parameters. 

Throughout this paper, we made several simplifying assumptions that could be lifted in future work. For instance, one could use the FFTLog algorithm to go beyond the  flat sky approximation used here~\citep{Assassi2017EfficientStatistics}. One could also improve the theoretical modelling by taking more orders of the perturbative expansion into account in order to push $k_{\text{max}}$ higher and closer to the more optimistic value mentioned above (see section~\ref{ssec:OptimisticForecast} and Tables~\ref{tab:Cosmokmax}, \ref{tab:PNGkmax}). 

\section*{Acknowledgements}
We would like to thank Muntazir Abidi and Blake Sherwin for useful discussions.
OL was funded by a Cambridge Trust European Scholarship and an STFC studentship. TB acknowledges support from the Cambridge Center for Theoretical Cosmology through a Stephen Hawking Advanced Fellowship. JF and PS acknowledge funding from STFC Consolidated Grant ST/P000673/1. DiRAC supermoputer resources in Cambridge were funded by BEIS capital grants ST/J005673/1 and STFC grants ST/H008586/1, ST/K00333X/1.

\bibliographystyle{mnras}
\bibliography{references}

\appendix

\section{CMB lensing Window function}
\label{sec:App1}
As outlined in section~\ref{ssec:obsSpectra}, we include the time dependency of the matter fields into the window functions when projecting (see~(\ref{eq:ProjInt})). While we ignore the time dependence in the thin tomographic bins of galaxy clustering, it cannot be neglected for CMB-lensing. This leads to different projection kernels for each perturbative order of the convergence field. In this work we are interested in the window functions at the first two orders, but the formalism outlined here is fully general.

The lensing window function in real space without the time evolution is given by
\[ W_\kappa = c \cdot (\chi_s/2+\chi)(\chi_s/2-\chi)/\chi_s \, \theta(\chi_s/2+\chi) \theta(\chi_s/2-\chi) \]
where $\chi_s$ is the comoving distance of the surface of last scattering and we centered the coordinate system at $\chi_s/2$. In an Einstein-deSitter Universe, the time evolution can be separated from the density contrast at all orders and the $n^{th}$ order perturbation comes with a factor $D^n/a$.

Our strategy is to approximate the time evolution with polynomials since they allow us to analytically integrate the combined window function. We do this by fitting a sixth order polynomial centered around $\chi_s/2$ to the growth factor using the least square method and weights $1/D$. The sixth order approximation leads to a relative error of less than $10^{-4}$ for the second order. We refer to Fig.~\ref{fig:LensingWindow} to see the behavior of linear and quadratic lensing function. 

The Fourier transform of the lensing window function can be expressed in terms of spherical Bessel functions, which is why we review them quickly in subsection~\ref{ssec:j_l} before we actually Fourier transform the polynomial real-space approximation to the window function in subsection~\ref{ssec:CMBint}. Finally, due to the slow decay of the window functions, one has to be careful with the one dimensional integral in the separation of the bispectrum~(\ref{eq:BS2D}). We outline in~\ref{ssec:cmb_proj} how one can make sense of supposedly divergent one dimensional integrals and separate the bispectrum.

\subsection{Spherical Bessel functions}
\label{ssec:j_l}
The spherical Bessel functions can be defined as
\[ j_n(z) = (-1)^n z^n \left(\frac{1}{z} \frac{d}{dz} \right)^n \frac{\sin(z) }{z} .\]
The first five are
\begin{equation}
    \begin{split}
j_0(z) =& \frac{\sin(z) }{z} \\
j_1(z) =& \frac{\sin(z) }{z^2}-\frac{\cos(z)}{z} \\
j_2(z) =&  \left(\frac{3}{z^2} - 1\right) \frac{\sin(z) }{z} -3 \, \frac{\cos(z)}{z^2} \\
j_3(z) =&  \left(1-\frac{15}{z^2} \right) \frac{ \cos(z) }{z} + \left(\frac{15}{z^3} - \frac{6}{z}\right) \frac{\sin(z)}{z}\\
j_4(z) =& \left(\frac{10}{z} -\frac{105}{z^3}\right) \frac{\cos(z)}{z} + \left( 1 - \frac{45}{z^2} + \frac{105}{z^4} \right)  \frac{ \sin(z)}{z}    .
    \end{split}
\end{equation}
For numerical accuracy, it is useful to utilize the low $z$ approximation
\begin{equation}
    j_l(z) \simeq \frac{z^l}{(2l+1)!!}\left(1- \frac{z^2}{6+4l} \right)
\end{equation} 
for $z\ll 1$.

\subsection{Fourier transform of monomial times CMB lensing window}
\label{ssec:CMBint}
 The Fourier transform of each monomial can be written as 
\begin{equation}
    \begin{split}
        \int_{-\chi_s/2}^{\chi_s/2} \frac{d\chi}{\chi_s}  \chi^n  (\chi_s/2&-\chi) (\chi_s/2+\chi) \exp[i \chi k]  = \\
        =&\frac{1}{2} \left( \frac{\chi_s}{2} \right)^{n+2} \int_{-1}^1 dx \, (1-x^2) x^n \exp[i x q ]  
    \end{split}
    \label{eq:CMBpre}
\end{equation}
where we substituted $q = \chi_s k/2$. This equation is then solved in terms of spherical Bessel functions as follows:
\begin{equation}
\begin{split}
\int d\chi  &W_\kappa \exp[i\chi k]\, \chi^n= \\
&= \begin{cases} 
       2 \left( \frac{\chi_s}{2} \right)^2 \frac{j_1(q)}{q}  & n=0 \\
      2 i \left(\frac{\chi_s}{2} \right)^3 \frac{j_2(q)}{q} & n=1 \\
       2 \left(\frac{\chi_s}{2} \right)^4 \left(\frac{j_2(q)}{q^2}-\frac{j_3(q)}{q}\right) & n=2 \\
       2 i \left(\frac{\chi_s}{2} \right)^5 \left(3 \, \frac{j_3(q)}{q^2}-\frac{j_4(q)}{q}\right) & n=3 \\
       2 \left(\frac{\chi_s}{2} \right)^6 \left( \frac{j_5(q)}{q}-6 \frac{j_4(q)}{q^2} + 3 \frac{j_3(q)}{q^3} \right)  & n=4 \\
       2 i \left(\frac{\chi_s}{2} \right)^7 \left( -\frac{j_6(q)}{q}- 10 \frac{j_5(q)}{q^2} + 15 \frac{j_4(q)}{q^3} \right) & n=5
   \end{cases}
  \end{split}
\label{eq:SphBessExp}
\end{equation}
The window function in Fourier space is then given by a weighted sum of the relevant monomials. 

\subsection{Integral identities for spherical Bessel functions}
\label{ssec:cmb_proj}
Having obtained an analytical expression for the lensing window function at all orders, we now discuss the projection integral of the bispectrum~(\ref{eq:BS2D}). Due to the slow decay of the lensing window function, the angular dependency leads to divergent 1D integrals when separating
\[ \mu(k_1,k_2) = \frac{k_3^2 - k_1^2 - k_2^2 }{2 k_1 k_2} .\]
Terms that involve the power spectrum decay quickly enough, so that they can be solved via standard Fast Fourier Transform (FFT) methods. The other lensing terms either suffer from a very slow decay or even diverge. However, using the the Bessel function's integral representation in terms of a Legendre polynomial $\mathcal{P}_l$
\begin{equation}
j_l(kr)=\frac{(-i)^l}{2}\int_{-1}^{1} d\mu \ \mathcal{P}_l(\mu)e^{i k r \mu}
\end{equation}
allows to compute Fourier transforms of spherical Bessel functions with polynomial coefficients for $n\geq 0$ as follows\footnote{For n=0, that is simply the inverse Fourier transform and one recovers the real space window function.}
\begin{equation}
    \begin{split}
        & \int dk  \, k^n j_l(kr) \exp [- i kx] \\
        &= \int_{-1}^1 d\mu \, \mathcal{P}_l(\mu) \int dk \, k^n \exp[i k (r \mu - x)] \frac{(-i)^l}{2} \\
        &= \int_{-1}^1 d\mu \, \mathcal{P}_l(\mu) \int dk \, \frac{1}{(i r)^n } \frac{\partial^n}{\partial \mu^n} \exp[i k (r \mu - x)] \frac{(-i)^l}{2} \\
        &= \int_{-1}^1 d\mu \, \mathcal{P}_l(\mu) \int d\tilde{k} \, \frac{1}{(i r)^n r } \frac{\partial^n}{\partial \mu^n} \exp[i \tilde{k} (\mu - x/r)] \frac{(-i)^l}{2} \\
        &= \int_{-1}^1 d\mu \, \mathcal{P}_l(\mu) \frac{1}{(i r)^n r } \frac{\partial^n}{\partial \mu^n} \int d\tilde{k} \,  \exp[i \tilde{k} (\mu - x/r)] \frac{(-i)^l}{2} \\
        &= \int_{-1}^1 d\mu \, \mathcal{P}_l(\mu) \frac{1}{(i r)^n r } \frac{\partial^n}{\partial \mu^n} (2 \pi \delta^{D}(\mu - x/r)) \frac{(-i)^l}{2} \\
        &= \int_{-1}^1 d\mu \, \mathcal{P}_l(\mu) \frac{1}{(i r)^n r } \pi (-i)^l \frac{\partial^n}{\partial \mu^n} \delta^D(\mu - x/r) \\
        &= \int_{-1}^1 d\mu \, \frac{\pi (-i)^{l+n}}{r^{n+1} }    \delta^D(\mu - x/r) (-1)^n \frac{\partial^n}{\partial \mu^n}\mathcal{P}_l(\mu)  \\
        &= \begin{cases} 
      \frac{\pi }{r^{n+1}} (-1)^n (-i)^{n+l}  \left. \frac{\partial^n}{\partial \mu^n}\mathcal{P}_l(\mu) \right|_{\mu=\frac{x}{r}} & \, \text{if } |x/r|<1\\
      0 & \text{otherwise}   \: .
   \end{cases}
    \end{split}
\end{equation}
In summary, the relations presented here allow us to perform the Fourier transformations analytically that cannot be solved with the FFT method. This allows to separate the bispectrum projection integral and to use caching which makes it possible to include all clustering-lensing cross-spectra into the analysis.

\section{Power Spectrum Signal-to-Noise}
\label{sec:AppPS}
The analytical results for the power spectrum signal-to-noise (SN) with Gaussian covariance are summarized as follows: The total SN scales quadratically in $k_{\text{max} }/k_f $. Moreover, the total SN scales linearly with the number of included auto-spectra assuming all cross-spectra are included in the signal too. These two features are independent of shot-noise, redshift-space distortions and only depend on the assumed Gaussian covariance structure. 

In this Appendix we first show the scaling with $k_{\text{max}}/k_f$ for a single auto-spectra in subsection~\ref{ssec:PSkmaxScal} and then demonstrate the scaling with the number of bins in subsection~\ref{ssec:PSnScal} which implies the $k_{\text{max} }/k_f$ scaling outlined before holds there too.

\subsection{Single spectra}
\label{ssec:PSkmaxScal}
Assuming a Gaussian covariance structure and using~(\ref{eq:Cov-PS}), we can compute the ($k_{\text{max}}/k_f $) scaling of the SN of a single auto-spectrum as
\begin{equation}
    \begin{split}
        \text{SN}_{1 \text{auto} } &= \sum_{ij} P(k_i) C(k_i,k_j)^{-1} P(k_j)         \\
        &= \sum_i \frac{P(k_i)^2}{ \frac{k_f^2}{2 \pi k_i \Delta k} \, 2 P(k_i)^2 } = \sum_i \frac{\pi k_i \Delta k}{k_f^2} \\
        &= \frac{\pi \Delta k}{k_f^2} \sum_{i=0}^{n-1} ( (i+0.5) \cdot \Delta k + k_f)  \\
        &= \frac{\pi \Delta k}{k_f^2}  \left( n k_f + \frac{n^2}{2} \Delta k  \right) =\frac{\pi}{2}\left[ \left( \frac{k_{\text{max} }}{k_f}\right)^2 -1 \right]  .
    \end{split}
    \label{eq:SNscalingKmax}
\end{equation}
Here, we start the binning in k-space at the fundamental frequency $k_f$ and use $\Delta k$ as k-binning. So the relation between the number of k-bins and $k_{\text{max} }$ is given by $n = \left \lfloor \frac{k_{\text{max} }-k_f}{\Delta k} \right \rfloor $. Ignoring this discretization effect, we see that SN scales quadratically in $k_{\text{max} }/k_f$.

Cross-power spectra $P_{X Y}$  in contrast have a different covariance structure
\[ C_{XY, XY}(k_i, k_j) \propto \delta_{ij}^K \big (P_{X Y}^2(k_i) + P_{X X}(k_i) P_{Y Y}(k_i) \big ) \]
Accordingly there is no analytic result and the SN is determined by the ratio $P_{X Y}^2/(P_X P_Y)$.

\subsection{Combining spectra}
\label{ssec:PSnScal}
As different auto-spectra are correlated, their individual signal-to-noises do not simply add up. The cross-covariance of two auto-spectra has the form~(\ref{eq:Cov-PS})
\[ C_{XX, YY}(k_i,k_j) \propto \delta_{ij}^K P_{XY}^2(k) .\]
Including the appropriate $P_{XY}$ into the analysis allows to remove the correlation between the auto-spectra. We demonstrate this explicitly for two and three different auto-spectra and the general case follows from induction.

We start with two galaxy clustering bins. Due to homogeneity, we know that different scales are uncorrelated, so the result from the previous subsection applies, and we only have to investigate the effect of the two bins for one particular $k$. The scaling with $k_{\text{max} }$ can then be derived from above. The signal vector is then given by $S = 
    \begin{bmatrix}
     P_0      &   P_0     &  P_1 
\end{bmatrix}
$
where $P_0$ is the auto-power spectrum and $P_1$ the cross-power spectrum. The covariance is given by
\begin{equation} C_P = 
    \begin{bmatrix}
     2 P_0^2      &   2 P_1^2     &  2 P_0 P_1 \\
     2 P_1^2      &   2 P_0^2       & 2 P_0 P_1 \\
    2 P_0 P_1        &  2 P_0 P_1      & P_0^2 + P_1^2
\end{bmatrix}.
\end{equation}
where we dropped the mode counting prefactor $\frac{k_f^2}{2 \pi k_i \Delta k}$. One sees that the $P_1^2$ term prohibits a simple adding of the SN's from bins one and two. However, adding $P_1$ decorrelates the two bins effectively and one obtains 
\[ \text{SN}_{ \delta_{\text{g}} \delta_{\text{g}}}(2\, \text{bin}) = S C^{-1} S =  1 =  2 \: \text{SN}_{ \delta_{\text{g}} \delta_{\text{g}}}(1 \, \text{bin}).\]
This effect relies not on the two galaxy clustering bins having the same autospectra. Next, we include CMB-lensing $\langle \kappa \kappa \rangle' = K$ and the required clustering-lensing cross-correlations ($\langle \kappa \delta_i \rangle' = L_i$) into the signal vector.
\begin{equation} S = 
    \begin{bmatrix}
     P_0      &   P_0     &  P_1  & L_0 & L_1 & K
\end{bmatrix}.
\end{equation}
We have already seen the covariance of $S = \begin{bmatrix} P_0 & P_0 & P_1 \end{bmatrix}$, so it remains to compute the covariance of $S_2 = \begin{bmatrix} L_0 & L_1 & K \end{bmatrix}$ and their cross-covariance. They are given by:
\begin{equation}
    C_L = 
    \begin{bmatrix} 
        P_0 K + L_0^2 & P_1 K + L_0 L_1 & 2 L_0 K \\
        P_1 K + L_0 L_1 & P_0 K + L_1^2 & 2 L_1 K \\
        2 L_0K & 2 L_1 K & 2 K^2
    \end{bmatrix}
\end{equation}{}
\begin{equation}
    C_{PL} = 
    \begin{bmatrix} 
        2 P_0 L_0 & 2 P_1 L_0 & 2 L_0^2 \\
        2 P_1 L_1 & 2 P_0 L_1 & 2 L_1^2 \\
        P_0 L_1 + P_1 L_0 & P_1 L_1 + P_0 L_0 & 2 L_0 L_1
    \end{bmatrix}
\end{equation}{}
In total, we obtain:
\begin{equation} 
 C = \begin{bmatrix} C_P & C_{PL} \\ C_{PL}^T & C_L   \end{bmatrix} 
 \end{equation}
and again, adding the cross-correlations allows adding the auto-signal-to-noises directly
\[ \text{SN}_{\text{tot} } = S C^{-1} S = 3 \: \text{SN}_{ \delta_{\text{g}} \delta_{\text{g}}}(1 \, \text{bin}).\] 
In general
\[ \text{SN} = \text{\# \text{auto-spectra} } \:\cdot \: \frac{\pi}{2}\left[ \left( \frac{k_{\text{max} }}{k_f}\right)^2 -1 \right]  .\]
This result illustrates the problem of small scale information being projected onto larger scales. Assuming a fixed survey volume, the total SN increases linearly with the number of bins despite the fact that $k_{\text{max} }$ is fixed. The infinite information comes from small scales that pollute the large scales increasingly as the kernel shrinks in real space.

In addition, it turns out that adding cross-spectra as signal to an analysis that also contains auto-spectra does not add any information beyond 'decorrelating' the included auto-spectra. I.e. adding the clustering-lensing cross-spectra does not improve the clustering or lensing SN, it only benefits a joint analysis where it breaks the correlation. 

\section{Restricting maximal correlation length}
\label{App:MCL}

\begin{table}
        \caption{The maximal correlation lengths needed to recover 99\% of the Fisher Information for all cosmological parameters when including all cross-bin-correlation. Those results are independent of the FoG model. All entries are in \Mpc.}
    \label{tab:mcl}
    \begin{tabular}{ l c c c c c }
    \hline
    & $ \delta_{\text{g}} \delta_{\text{g}}$ & PS & $\delta_{\text{g}} \delta_{\text{g}} \delta_{\text{g}}$ & gal & total \\ [0.5ex]
    \hline
    n=2 & 590 & 590 & 590 & 590 & 590  \\
    n=4 & 295 & 295 & 295 & 295 & 295  \\
    n=6 & 295 & 295 & 295 & 295 & 295  \\
    n=8 & 221 & 221 & 221 & 221 & 221  \\
    n=10 & 177 & 177 & 177 & 177 & 177  \\
    n=12 & 147 & 147 & 196 & 196 & 196  \\
    n=16 & 110 & 110 & 184 & 184 & 184  \\
    \hline
    \end{tabular}   
\end{table}
\begin{figure*}
\includegraphics[width=\textwidth]{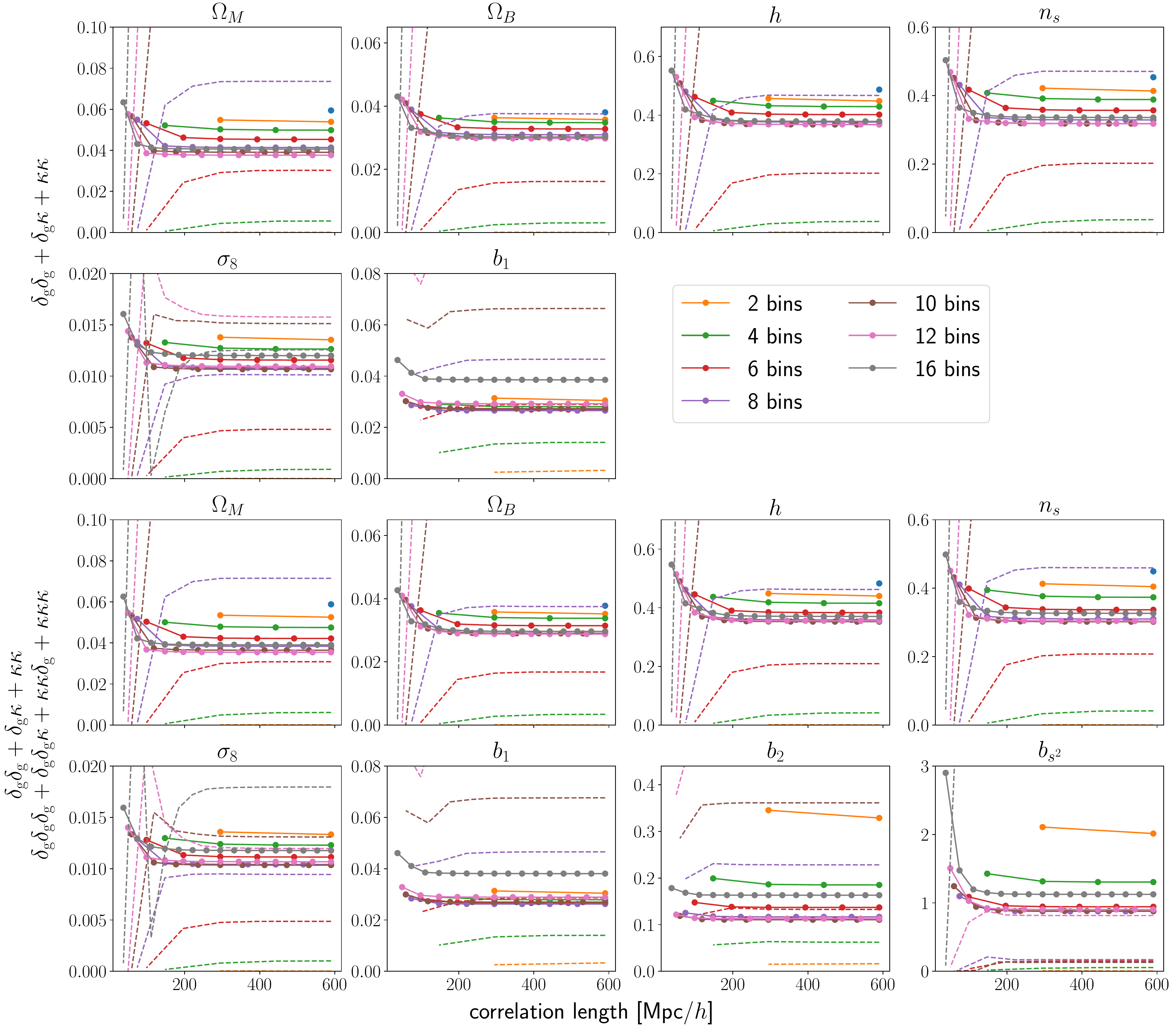}
\caption{Error bars (solid lines) and fives times the biases (dashed lines) for joint clustering-lensing spectra as a function of the maximal correlation length (see~(\ref{eq:mcl})) using FoG model 2. The intersection of dashed and solid lines, marks the point where the relative bias becomes $20\%$. The top two rows correspond to a power spectrum analysis and the bottom two rows show an analysis with all power spectra and bispectra. Each subplot shows the scaling for a different parameter.}
\label{fig:mclClusteringLensing}
\end{figure*}
The number of cross-bispectra configuration is prohibitively large if one wants to perform an MCMC fit or estimate empirical covariance matrices from simulations. This is why several authors have studied compression techniques~\citep{Fergusson2012RapidStructure, Schmittfull2015NearStructure,Eggemeier2017CosmologyBAO, Byun2017TowardsStatistics,Gualdi2018}. 
In this Appendix, we reduce the number of galaxy cross-spectra by introducing a maximal correlation length for all spectra that contain two or more galaxy fields ($\langle  \delta_{\text{g}} \delta_{\text{g}} \rangle$, $\langle  \delta_{\text{g}} \delta_{\text{g}} \kappa \rangle$ and $\langle \delta_{\text{g}} \delta_{\text{g}} \delta_{\text{g}} \rangle$). We define the maximum correlation length as
\begin{equation}
\text{mcl} = d/n \cdot \left( \max(\bmath{\chi}) - \min(\bmath{\chi}) \right) \label{eq:mcl}
\end{equation}
where $d$ is the depth of the survey, $n$ the total number of tomographic bins, and $\bmath{\chi}$ labels the galaxy bins with larger values assigned to bins that are further away from the observer. Restricting the maximal correlation length reduces the power spectrum scaling from $n^2$ to $n\cdot \text{mcl}$ and for the bispectrum $n^3$ to $n\cdot \text{mcl}^2$.

We summarize the maximal correlation length needed in order to recover at least $99\%$ of the Fisher information of the analysis with all cross-bin correlations in Table~\ref{tab:mcl}. We find that the maximum correlation length is independent of the FoG model used. Since all combinations of spectra show roughly the same maximal correlation length needed (see Table~\ref{tab:mcl}), we show in Figure~\ref{fig:mclClusteringLensing} only the scaling for two spectra in full detail. One observes that restricting the maximal correlation lengths leads not better error-bias trade-offs than the ones we discussed in subsection~\ref{ssec:tradeoff}. This is due to the fact that biases arise from inaccurate modelling of radial modes, which is precisely the type of information one excludes when decreasing the correlation length.

\end{document}